\documentclass[journal,twocolumn]{IEEEtran}
\IEEEoverridecommandlockouts                              

\usepackage[noadjust]{cite}
\usepackage[utf8]{inputenc} 
\usepackage[T1]{fontenc} 
\usepackage{hyperref}       
\usepackage{url}            
\usepackage{booktabs}      
\usepackage{amsfonts}      
\usepackage{nicefrac}      
\usepackage{microtype}    
\usepackage{algorithmic}
\usepackage[ruled,vlined]{algorithm2e}
\usepackage{amsmath}
\usepackage{graphicx}
\usepackage{multicol}
\usepackage[font=footnotesize, labelfont=bf]{caption}
\DeclareMathOperator{\Tr}{Tr}

\usepackage{dblfloatfix}

\usepackage{subcaption,graphicx}

\usepackage{hyperref}
\hypersetup{colorlinks=true,linkcolor=blue,urlcolor=blue}

\usepackage[]{algorithm2e}
\SetKwInput{KwInput}{Input \hspace*{0.3em}}
\SetKwInput{KwOutput}{Output}

\usepackage{amsthm}
\theoremstyle{remark}

\usepackage{tikz}

\usepackage{amssymb}
\usepackage{pgfplots}
\usepackage{pgfmath}
\usepgfplotslibrary{patchplots}
\usetikzlibrary{patterns, positioning, arrows}
\usepackage[shortlabels]{enumitem}
\newtheorem*{remark}{Remark}

\pgfmathsetmacro\sprayRadius{.75pt}
\pgfmathsetmacro\sprayPeriod{.8cm}

\usepackage{textcomp}
\usepackage{lipsum}

\newcommand\copyrighttext{%
  \footnotesize \textcopyright 2022 IEEE. Personal use of this material is permitted.
  Permission from IEEE must be obtained for all other uses in any current or future
  media, including reprinting/republishing this material for advertising or promotional
  purposes, creating new collective works, for resale or redistribution to servers or
  lists, or reuse of any copyrighted component of this work in other works.
  DOI: \href{<http://tex.stackexchange.com>}{10.1109/TNNLS.2022.3172108}}
\newcommand\copyrightnotice{%
\begin{tikzpicture}[remember picture,overlay]
\node[anchor=south,yshift=3pt] at (current page.south) {\fbox{\parbox{\dimexpr\textwidth-\fboxsep-\fboxrule\relax}{\copyrighttext}}};
\end{tikzpicture}%
}

\title{
Tensor-CSPNet: A Novel Geometric Deep Learning Framework for Motor Imagery Classification
}

\author{
Ce~Ju and Cuntai~Guan~\IEEEmembership{Fellow, IEEE}
\thanks{Ce Ju and Cuntai Guan are with the S-Lab, Nanyang Technological University, 50 Nanyang Avenue, Singapore (emails: \{juce0001,ctguan\}@ntu.edu.sg). 
}
}

\begin{document}

\maketitle

\copyrightnotice

\begin{abstract}
Deep learning (DL) has been widely investigated in a vast majority of applications in electroencephalography (EEG)-based brain-computer interfaces (BCIs), especially for motor imagery (MI) classification in the past five years.
The mainstream DL methodology for the MI-EEG classification exploits the temporospatial patterns of EEG signals using convolutional neural networks (CNNs), which have remarkably succeeded in visual images. 
However, since the statistical characteristics of visual images depart radically from EEG signals, a natural question arises whether an alternative network architecture exists apart from CNNs.
To address this question, we propose a novel geometric deep learning (GDL) framework called Tensor-CSPNet, which characterizes spatial covariance matrices derived from EEG signals on symmetric positive definite (SPD) manifolds and fully captures the temporospatiofrequency patterns using existing deep neural networks on SPD manifolds, integrating with experiences from many successful MI-EEG classifiers to optimize the framework. 
In the experiments, Tensor-CSPNet attains or slightly outperforms the current state-of-the-art performance on the cross-validation and holdout scenarios in two commonly-used MI-EEG datasets. 
Moreover, the visualization and interpretability analyses also exhibit the validity of Tensor-CSPNet for the MI-EEG classification. 
To conclude, in this study, we provide a feasible answer to the question by generalizing the DL methodologies on SPD manifolds, which indicates the start of a specific GDL methodology for the MI-EEG classification. 
\end{abstract}


\IEEEpeerreviewmaketitle

\section{INTRODUCTION}
A brain-computer interface (BCI) is a direct communication pathway between a user’s brain and an external device by measuring and analyzing the behaviorally relevant information of brain activities.~\cite{wolpaw2002brain}
The non-invasive electroencephalogram (EEG)-based BCI is one of the most common BCIs, employing portable, non-invasive electrodes on the scalp for instantaneously measuring electrical changes in neurons. 
It allows brain-derived communication between patients with amyotrophic lateral sclerosis and motor control restoration after stroking and spinal cord injury.~\cite{machado2010eeg}
However, decoding mental states from EEG-based BCI is challenging for various reasons, such as low signal-to-noise ratio (SNR), artifacts, and high inter/intra-subject variabilities (a.k.a, nonstationarity changes) in EEG signals.~\cite{blankertz2007optimizing} 

In the paradigm of traditional EEG analysis, spatial patterns of EEG signals are crafted by a preprocessing algorithm to have more strong discrimination between mental states and afterward classified using machine-learning classifiers, such as the support vector machine (SVM) and linear discriminant analysis (LDA).~\cite{lotte2018review}  
Many spatial filterings such as common spatial pattern (CSP) and its variants~\cite{koles1990spatial,ang2008filter,lu2010regularized,lotte2010regularizing} are widely used as such preprocessing algorithms to increase the SNR of signals and, therefore, enhance oscillatory brain electrical activities before feature extraction. 
However, the validness of the analysis is limited to the capacity of feature extraction for complex event-related and event-unrelated (resting state) neural oscillations.~\cite{lawhern2018eegnet}

To remedy this limitation, the architecture of CNNs has been broadly adopted as an emerging tool in BCIs~\cite{schirrmeister2017deep,sakhavi2018learning,lawhern2018eegnet,bang2021spatio,mane2021fbcnet,stieger2021benefits}. 
Technically, the scheme of these CNN classifiers is designed to automatically capture the temporospatiofrequency features of neural signals in end-to-end learning without the experience of human engineers. 
It has been proven effective for the MI-EEG classification in literature~\cite{lotte2018review,mane2021fbcnet}.
Compared with the previous non-DL approaches, CNN is making significant advances in the incredible power of representation with multiple levels of abstraction, end-to-end learning, and causal contributions of patterns on brain topography.~\cite{schirrmeister2017deep} 
However, the essential difference in the underlying structure between images and EEG signals discernibly weakens the feature expression of CNNs in BCI tasks. 
Specifically, several prior assumptions in computer vision require the underlying structure of visual images to be \emph{stationary}, \emph{translation invariant}, \emph{translation equivariant}, and \emph{stable} with respect to local deformations, conceptually characterized as the Euclidean nature, which enables CNNs to effectively extract local features from local statistics.~\cite{bruna2013invariant,lecun2015deep,bronstein2017geometric}
In contrast, the underlying structure of EEG signals might not embody the Euclidean nature according to electrophysiological studies and (nonlinear) dynamical neuroscience~\cite{moehlis2008dynamical,izhikevich2007dynamical}.
To illustrate, a prominent example is that EEG signals are non-stationary, and their local statistics are variant to the location of spatially distributed regions. 
Consequently, a natural question arises whether an alternative network architecture exists apart from CNNs for efficient feature extraction in the MI-EEG classification.

In the history of the MI-EEG classification study, such an alternative discipline has been raised, which uses a graph convolutional neural network to learn the graph signal representations of EEG rhythmic components~\cite{jang2018eeg}. 
Apart from their approach, in this study, we set off a novel discipline in terms of spatial covariance matrice (SCM) derived from EEG signals, which is the inherent correlation between neighbor channels, a second statistics in the spatial domain, and have been developed in CSP for over 30 years~\cite{koles1990spatial,muller1999designing}. 
We aim to formulate SCMs using Riemannian geometry for an in-depth analysis of the non-Euclidean nature as many existing Riemannian geometry-based modelings in engineering disciplines such as diffusion tensor imaging and geometric mechanics~\cite{pennec2006riemannian,pennec2020manifold,libermann2012symplectic}. 
The Riemannian-based BCI classifier that characterizes EEG signals using the geometric information of SCMs emerged about a decade ago.~\cite{barachant2011multiclass,congedo2013new,yger2016riemannian}
It has gained growing interest from the BCI community, and various follow-ups were proposed to optimize the structure~\cite{xie2016motor,congedo2017riemannian,sabbagh2019manifold}.
Technically, SCMs derived from EEG signals are inherently symmetric positive definite (SPD).
The space of SPD matrices is formulated as a Riemannian manifold called the SPD manifold, provided with a specific metric. 
Then, the geodesic distance on SPD manifolds between two SCMs is encoded as a high-level feature for the machine-learning classifier.

The most fruitful part of the Riemannian-based BCI classifier is the conceptual importance of using SPD manifolds to characterize EEG signals. 
However, there are many practical drawbacks to the Riemannian-based BCI classifier. 
Firstly, hand-crafted feature extraction is outdated and inefficient in complex scenarios such as feature expression for non-homogeneous BCI sensor data. 
Secondly, the neurophysiological interpretation of existing hand-crafted features such as geodesic distance on SPD manifolds has not yet been fully understood~\cite{kobler2021interpretation}. 
To cope with these practical drawbacks, more recently, a novel classifier on SPD manifolds~\cite{ju2020federated} is probably a promising solution, which investigates the low-level features of SCMs for the EEG classification using SPDNet~\cite{huang2017riemannian}, an existing Riemannian-based network architecture, to capture the spatial patterns of EEG rhythmic components.
Architecture SPDNet is a DL architecture that preserves the SPD structure of matrices across layers and exhibits competitive performance compared with the current state-of-the-art approaches using CNNs on an increasing number of computer vision tasks.~\cite{huang2015log,huang2017deep}
Its perspective generally originates from an emerging subfield geometric deep learning (GDL)~\cite{bronstein2017geometric}, which aims to generalize the DL models to the non-Euclidean domain as graphs and manifolds.

In this paper, we propose a novel GDL framework, \emph{Tensor-CSPNet}, that generalizes the DL methodology for the MI-EEG classification. 
To this end, we build a network architecture upon the principle that largely exploits the temporospatiofrequency patterns of EEG signals. 
Each structure in our architecture aims to capture features from either of the temporospatiofrequency domains.
Firstly, tensor stacking segments EEG signals and stacks them into the temporospatiofrequency tensors according to hand-crafted technical and neurological experience.
Each tensor is in an SPD-matrix representation that encodes the inherent correlation between neighbor channels with respect to the time and frequency information, which is also a rough estimation of the brain connectivity between spatially segregated areas~\cite{8786184}. 
Secondly, the spatial patterns and temporal dynamics behind EEG signals are extracted by deep neural networks on SPD manifolds and CNNs on the tangent space sequentially and respectively. 
Significantly, the combination of the depthwise BiMap layer and ReEig consists of a nonlinear spatial filter that enhances the feature expression of spatial patterns. 
Finally, the classification stage classifies the extracted temporospatiofrequency patterns using fully-connected neural networks.

In the experiments, Tensor-CSPNet is investigated on several motor imagery (MI) tasks of EEG-based BCIs, including stationary and non-stationary scenarios.
Typically, an MI task refers to an experiment where the individual mentally simulates a physical action. 
In neurophysiology, since MI of motor actions produces replicable and discriminable patterns (i.e., synchronization/desynchronization) over the primary sensory and motor areas, the signals are discernible to be a classification task.~\cite{pfurtscheller2001motor,neuper2005imagery}
In addition, the visualization and interpretability analyses are conducted on two MI datasets to double verify the validity of Tensor-CSPNet. 
The remainder of this paper is organized as follows: 
Section~\ref{PRELIMINARY} introduces the paradigm of traditional EEG analysis and elaborates on the mathematical background of CSP and SPDNet. 
The subsequent section~\ref{Methodology} is the methodology for Tensor-CSPNet. 
The performance of Tensor-CSPNet is then compared on a broad set of experiments in Section~\ref{Experiments}, and cautious discussions of nonstationarity and contrasts with other mainstreams are in Section~\ref{Discussion}.
In Appendix, we discuss the prior assumptions for CNNs, a brief overview of SPD manifolds, the strategy of fixed-interval segmentation, and an ablation study on BCIC-IV-2a. 

\begin{figure*}[!h]
  \centering
  \includegraphics[width=18cm]{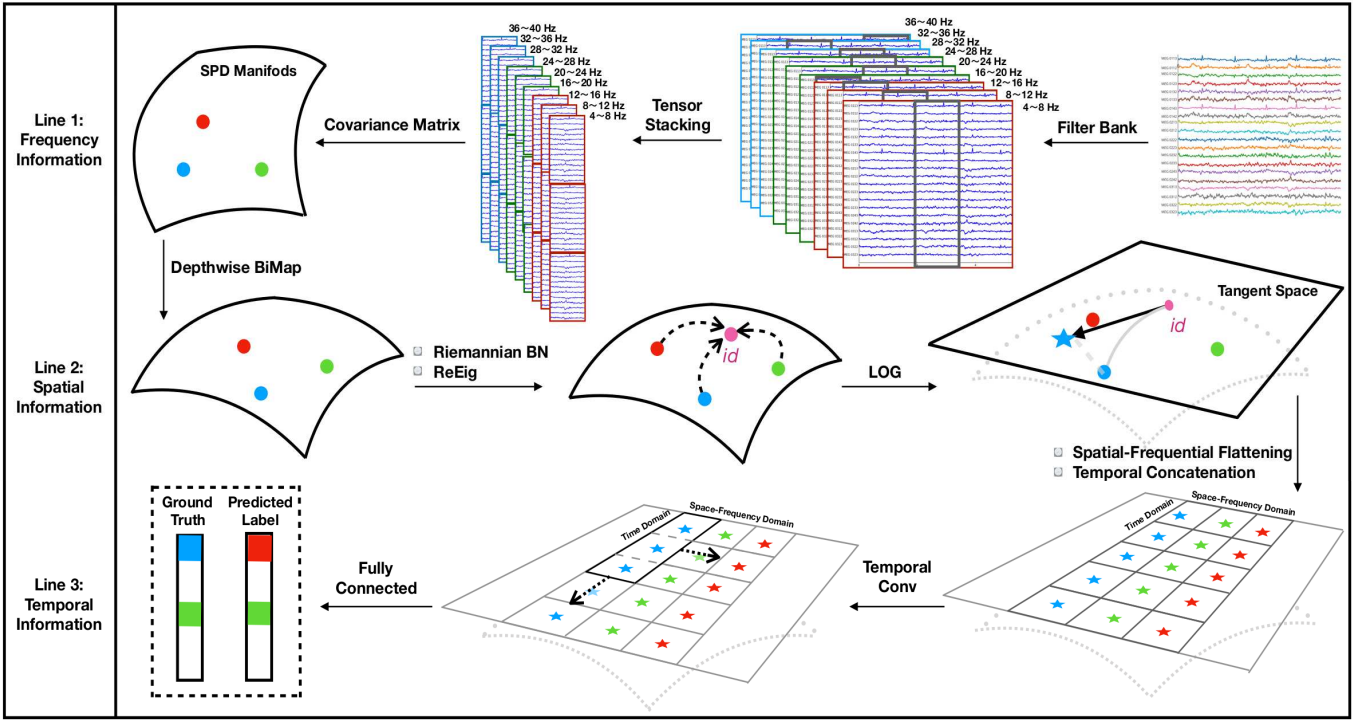}
  \caption{Illustration of Architecture of Tensor-CSPNet: 
The network architecture is built upon the principle that fully exploits the temporospatiofrequency patterns behind EEG signals. 
Hence, each structure in the architecture aims to capture features from either of the temporospatiofrequency domains.
In Line 1, EEG signals are segmented into the temporospatiofrequency tensors in the tensor stacking stage. Frequency information has been unfolded in this stage.  
In Line 2, the CSP stage is designed to capture spatial information from tensors using the depthwise BiMap layer, Riemannian BN layer, and the ReEig layer. 
In Line 3, we capture the temporal information on the tangent space using 2D CNNs. 
Fully connected neural networks with cross-entropy loss are used for the MI-EEG classification.
\label{architecture}
} 
\end{figure*}

\section{PRELIMINARY}\label{PRELIMINARY}
\subsection{Notations}
Let $ \mathcal{S}^n := \{ \, S \in \mathbb{R}^{n \times n} : S = S^\top \, \}$ be a set of $n \times n$ real symmetric matrices and $\mathcal{S}^n_{++} := \big\{ \, S \in \mathcal{S}^n : x^\top S x > 0, \forall x \in \mathbb{R}^n/\{0\} \, \big\}$ be a set of $n \times n$ real SPD matrices.
The \emph{Frobenius} inner product and norm on $m \times n $ matrices $A$ and $B$ are defined as $\langle \, A, B \, \rangle_{\mathcal{F}} := \Tr{(A^\top B)}$ and $||\,A\,||_{\mathcal{F}}^2 :=  \langle \, A, A \, \rangle_{\mathcal{F}}$ respectively.

\subsection{Paradigm of Traditional EEG Analysis}\label{pre:paradigm}
Let $X \in \mathbb{R}^{C \times T}$ be a short segment (trail) of EEG signals, where $C$ is the number of EEG channels (electrodes), and $T$ is the number of sampled points on epoch durations. 
This paper assumes that trail $X$ is already band-pass filtered, centered, and scaled. 
A linear classifier that predicts the label of trail $X$ is typically written as $f(X; \{w_i\}_{i=1}^N, \{\beta_i\}_{i=1}^N) = \sum_{i=1}^N \beta_i \log{(w_i X X^\top w_i^\top)} + \beta_0$, where $N$ is the number of spatial filters, $\{w_i\}_{i=1}^N \in \mathbb{R}^C$ are spatial filters and $\{\beta_i\}_{i=1}^N \in \mathbb{R}$ are biases. 
At many physiological and anatomical levels in the brain, the lognormal distributions are fundamental to structural and functional brain organization because the distribution of numerous parameters is strongly skewed with a heavy tail.~\cite{buzsaki2014log} 
Hence, the logarithm of the power/variance of the projected signal $w_i XX^\top w_i^\top$ is considered in the classifier. 
%

\subsection{Common Spatial Pattern}\label{pre:csp}
Let $\Sigma^{+}, \Sigma^{-} \in \mathbb{R}^{C \times C}$ be the estimates of covariance matrices of trails $\{X_i\}_{i=1}$ in a 2-class MI -EEG paradigm, i.e.,
$\Sigma^{c} = \frac{1}{|\mathcal{I}_c|} \cdot \sum_{i \in \mathcal{I}_c } X_c \cdot X_c^\top$,
where $\mathcal{I}_c$ $(c \in \{+, -\})$ is the set of indices of trails of one class. 
The CSP algorithm is given by a simultaneous diagonalization of covariance matrices $\Sigma^{+}$ and $\Sigma^{-}$ in two equations such as $W \cdot \Sigma^{+} \cdot W^\top = \Lambda^{+}$ and $W \cdot \Sigma^{-} \cdot W^\top = \Lambda^{-}$, where each column vector $w_i \in \text{col}(W)$ is a spatial filter in CSP. 
The diagonal matrices $\Lambda^{+}, \Lambda^{-} \in \mathbb{R}^{C \times C}$ hold an identity constraint, i.e., $\Lambda^{+} + \Lambda^{-} = I$. 
The problem of the above simultaneous diagonalization is mathematically equivalent to solve a generalized eigenvalue problem as follows: $(\Sigma^{+} w) = \lambda \cdot (\Sigma^{-} w)$.
Feature vectors $z:= \log{(w XX^\top w^\top)}$ consisting of eigenvectors $w \in$ col(W) from both ends of the eigenvalue spectrum are commonly used in the EEG analysis.

\subsection{Riemannian Batch Normalization}

Riemannian Batch Normalization (BN) is a generalization of the classic batch normalization on Riemannian manifolds.~\cite{brooks2019riemannian}. 
Formally, the weighted Riemannian barycenter on SPD manifolds $\text{Bar}_{\bold{w}} (\{\mathcal{B}\})$ utilizes the parallel transport $\Gamma$ to connect any sample $S^{i}$ with the identity matrix I$_d$ according to formulas $\Gamma_{\mathcal{B} \mapsto I_d} (S^{i}) := \mathcal{B}^{-\frac{1}{2}} \cdot S^{i} \cdot \mathcal{B}^{-\frac{1}{2}}$ and $\Gamma_{I_d \mapsto G} (S^{i}) := G^{\frac{1}{2}} \cdot S^{i} \cdot G^{\frac{1}{2}}$, where each mini batch $\mathcal{B}$ of SPD matrices $\{S^{i}\}_{i=1}$, the biasing parameter of $G$ acquired by the matrix backpropagation in the training is directly applied in the inference. The definition of the parallel transport and the weighted Riemannian barycenter refers to Appendix~\ref{appen:metric}.

\subsection{SPDNet}
Architecture SPDNet is a deep neural network architecture fed with SPD matrices that preserves the SPD structure of matrices across layers during non-linearly learning.~\cite{huang2017riemannian}
Analogous to convolutional neural networks, the basic layers in SPDNet are designed to include the following layers:
\begin{itemize}
\item BiMap: 
This layer transforms the covariance matrix $S$ using the bi-map operator $W\cdot S \cdot W^T$. 
Transformation matrix $W$ is required to be full row rank.
\item ReEig: This layer is analogous to ReEig in classical deep neural networks that introduces the non-linearity on SPD manifolds using $U \cdot \max{(\epsilon I, \Sigma)} \cdot U^T$, where singular value decomposition $S = U \cdot \Sigma \cdot U^T$, and $\epsilon$ is a rectification threshold and $I$ is an identity matrix.
\item LOG: This layer is to map elements on SPD manifolds on its tangent space using $U \cdot \log{(\Sigma)} \cdot U^T$, where singular value decomposition $S = U \cdot \Sigma \cdot U^T$.
\end{itemize}

\section{METHODOLOGY}\label{Methodology}

In this section, we propose a novel GDL framework \emph{Tensor-CSPNet} for non-invasive EEG-based BCIs, consisting of four stages: the tensor stacking stage, the common spatial pattern stage, the temporal convolutional stage, and the classification stage.
The architecture of Tensor-CSPNet is illustrated in Figure~\ref{architecture}. 
%

\subsection{Stage One: Tensor Stacking Stage}
In the first stage, EEG signals will be segmented into the temporospatiofrequency tensors concerning the theory of neurophysiology, electrophysiology, and signal processing.
%

\subsubsection{frequency Segmentation}
We use a well-known filter-bank technique in the EEG-BCI classification~\cite{ang2008filter} for frequency segmentation, which employs a bank of bandpass filters to decompose the raw oscillatory EEG signals into multiple frequency passbands using the causal Chebyshev Type II filter. 
%

\subsubsection{Temporal Segmentation}\label{TDD}
The temporal segmentation aims to divide EEG signals into small segments on the time domain with or without overlapping.  
Generally, the signals should be segmented according to the characteristics of EEG-based BCI tasks, for instance, dynamic changes in very short durations in many cognitive tasks.
For those signals that we are not familiar with their characteristics, we propose a fixed-interval segmentation strategy in Appdendix~\ref{appen:fixed} that EEG signals are initially subdivided into fixed short equal-length intervals without overlapping.
We require that the length of the time window $\omega$ (time resolution) is limited by Garbor's uncertainty principle~\cite{gabor1946theory} that time and frequency resolutions cannot be at a high level simultaneously. 
%

\subsubsection{Tensor Stacking}
After two segmentations, we stack elementary information cells to the four-dimensional temporospatiofrequency tensors $\tilde{X} \in \mathbb{R}^{W \times F \times C \times \omega}$, where $W$, $F$, $C$ and $\omega$ are the number of window slices, the number of filter banks, the number of channels and window length respectively. 
As a consequence, the input tensors of Tensor-CSPNet are spatial covariance matrices $S^{ij} := \tilde{X}[i, j, :, :] \cdot \tilde{X}[i, j, :, :]^\top$ for $i \in$ windows slices $W$ and $ j \in$ filter banks $F$.
The pseudocode of tensor stacking refers to Algorithm~\ref {alg}.

\begin{algorithm}
\SetAlgoLined
\KwInput{$X \in \mathbb{R}^{F \times C \times T}$, window length $\omega$, stride $s$, and padding value $p$)}
\KwOutput{$ \tilde{X} \in \mathbb{R}^{\lfloor \frac{T+2p -1}{s} +1 \rfloor \times F \times C \times \omega}$.}
\For{$i \gets 0$ $\KwTo \lfloor \frac{T+2p -1}{s} +1 \rfloor$ }{
\For{$ j \gets 0$ \KwTo F}{
    $\tilde{X}[i, j, :, :] \gets X[j, :, is: is + \omega]$ 
   }
}
\caption{Tensor Stacking Stage\label{alg}}
\end{algorithm}

\begin{remark}
1). The tensor stacking stage is a data preprocessing stage, which is not included in the network architecture. 

2). In frequency segmentation, we adopt a widely used portfolio of filter banks $\{ 4 \sim 8$ Hz, $8 \sim 12$ Hz, $\cdots$, $36 \sim 40$ Hz$\}$, which has exhibited the best competition results in the BCI Competition IV 2a~\footnote{\, The results of BCI competition IV can be found in the official website http://www.bbci.de/competition/iv/results/.} 
It is a well-known and widely used dataset in the MI-EEG classifiers.
There are also many pieces of literature to exploit different combinations of frequency ranges, but this one is the most straightforward. 
\end{remark}

\subsection{Stage Two: Common Spatial Pattern Stage}\label{stage2}
In the second stage, we modify and employ the architectures of SPDNet to capture the spatial patterns of EEG signals.
%
\subsubsection{Depthwise BiMap}
The BiMap layer in SPDNet~\cite{huang2017riemannian} will be first modified to the depthwise BiMap layer that does not take a channel-wise summation after the bi-multiplication and then employed in the CSP stage. 
Preserving the SPD structure, it will transform spatial covariance matrices in each channel by right-multiplying a full column-rank matrix $W$ and left-multiplying its transpose simultaneously, i.e., $W\cdot S^{ij} \cdot W^\top$.
%

\subsubsection{Riemannian BN}
Riemannian BN is used to de-correlate a batch of sample-based spatial covariance matrix estimation towards to an identity by $\Gamma_{\mathcal{B} \mapsto I_d} (S^{ij}) := \mathcal{B}^{-\frac{1}{2}} \cdot S^{ij} \cdot \mathcal{B}^{-\frac{1}{2}}$, which equalizes the variance in all directions and removes the batch effects. 
The behind statistical mechanism has been exhibited in CSP's variant, Regularized-CSP~\cite{lotte2010regularizing}, in which a shrinkage is performed towards the identity on the regularized estimate for each class as $\bar{S}^{ij} := (1-\gamma) \cdot \bar{S}^{ij} + \gamma \cdot I_d$, where $\bar{S}^{ij}$ is the regularized estimate of $S^{ij}$, $\gamma$ is a user-defined parameter.

\subsubsection{ReEig}
ReEig in SPDNet is used to consist of a nonlinear spatial filter. 
In contrast with the traditional EEG analysis in which spatial filters are linear, this layer enables Tensor-CSPNet to have a richer feature expression of spatial information.
\subsubsection{LOG}\label{tpl}
LOG in SPDNet is adopted to log-project the transformed SPD matrices onto the tangent space for the log-power/variance.
It is consistent with a step in the standard paradigm of EEG analysis, refer to Section~\ref{pre:paradigm}.

\subsection{Stage Three: Temporal Convolutional Stage}\label{TCL}
\begin{figure}[!h]
\centering
\includegraphics[width=7.5cm]{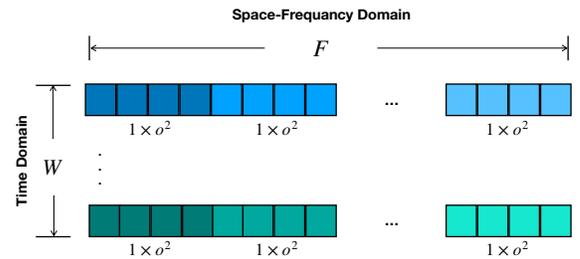}
\caption{Illustration of Temporal Convolutional Stage:
$F$ blocks of $1\times o^2$ rectangles flattened and $W$ lines concatenated. 
To illustrate, in the case of 5-CSPNet employed on MI-KU, $F = 9$, $o = 20$, and $W=5$.
Thus, each line is a $1\times3600$ flattened tensor, and the shape of the whole rectangle is $5\times 3600$. 
\label{TCN}
} 
\end{figure}

In this stage, we aim to capture temporal dynamics on tangent space using CNNs. 
We first flatten the outputs of the CSP stage on the frequency and space domains called the \emph{Spatial-frequency Flattening}.
Then, we concatenate the flattened tensors along the time domain called the \emph{Temporal Concatenation}.
After the spatial-frequency flattening and the temporal concatenation, the temporospatiofrequency tensor becomes a 2-dimensional tensor in $\mathbb{R}^{W \times (Fo^2)}$ without SPD structure anymore, where $F$ is the number of filter banks and $o$ is the output dimension of the CSP stage, as illustrated in Figure~\ref{TCN}. 
Finally, we use 2-dimensional (2D) CNN with $po^2$-width (p = 1 or $F$) and $q$-height ($1\leq q \leq W$) to capture the temporal dynamics of EEG signals. 

\begin{remark}
 
1). In Stage Three, the use of 2D CNN for extracting temporal dynamics is because, in principle, the tangent space at a point on a Riemannian manifold is a vector space isomorphic to Euclidean space with the same dimension and thus always flat. 
Hence, after LOG, the classification problem returns to one in the (flat) Euclidean domain. 
The geometric neural networks, developed to deal with problems in the curved space, are not necessary to apply. 

2). The width of 2D CNN is set at a multiple of $o^2$ because we hope to alleviate the influence of different spatial locations of EEG electrodes on the scalp. 
Two possible multiple $p = 1$ and $F$ mean that each frequency band in the portfolio can independently and equally contribute to the model performance.
\end{remark}

\subsection{Stage Four: Classification Stage and Loss Function}
In the final stage, single-layer or multi-layer neural networks are utilized for the final classification.
The loss function in our approach is cross-entropy for the sake of simplicity.

\section{Experiments}\label{Experiments}

\subsection{Evaluation Dataset}
We investigate the proposed approach on two MI datasets, including Korea University Dataset (MI-KU)~\cite{lee2019eeg} and the BCI Competition IV 2a (BCIC-IV-2a)~\cite{brunner2008bci}.

\subsubsection{Korea University Dataset (MI-KU)}
In the MI paradigm of the MI-KU dataset, 54 subjects performed a binary class MI task. 
The signals were collected with 62 Ag/AgCl electrodes where 20 electrodes in the motor cortex region were selected (FC-5/3/1/2/4/6, C-5/3/1/z/2/4/5, and CP-5/3/1/z/2/4/6) and recorded with a sampling rate of 1,000 Hz for our evaluation of each classifier. 
The MI-KU dataset has two sessions (S1 and S2), each with 200 trials per subject. 

\subsubsection{BCI Competition IV 2a (BCIC-IV-2a)}
BCIC-IV-2a is a cue-based BCI paradigm with four-class MI-EEG motor imagery tasks including left hand, right hand, feet, and tongue recorded in 22 Ag/AgCl EEG electrodes and three monopolar EOG channels with a sampling rate of 250 Hz from 9 subjects. 
The BCIC-IV-2a dataset has the training session (T), and the evaluation session (E) recorded on different days. 
Each subject performed six runs of 12 cue-based trials for each of the four classes in either training or evaluation sessions, yielding 288 trials per subject. 
%

\begin{table*}[!t]
\caption{Configurations of Temporal Segments: 
In this paper, there are three kinds of temporal segments without overlapping on \textbf{MI-KU}, and there are four kinds of temporal segments with overlapping on \textbf{BCIC-IV-2a} that are adopted from~\cite{zhang2018temporally}.
}
\centering 
\begin{tabular}{l l r }
\toprule
MI-KU &  Temporal Segments (sec.)  \\
\midrule
(a). 1-CSPNet & \{$1.0 \sim 3.5$\} \\
(b). 5-CSPNet & \{$1.0 \sim 1.5$,  $1.5 \sim 2.0$, $2.0 \sim 2.5$, $2.5 \sim 3.0$, $3.0 \sim 3.5$\} \\
(c). 10-CSPNet & \{$1.00 \sim 1.25$,  $1.25 \sim 1.50$, $1.50 \sim 1.75$, $1.75 \sim 2.00$,  $2.00 \sim 2.25$, $2.25 \sim 2.50$,  \\
&\,$2.50 \sim 2.75$, $2.75 \sim 3.00$, $3.00 \sim 3.25$, $3.25 \sim 3.50$\}\\
\toprule
BCIC-IV-2a  &  Temporal Segments (sec.)  \\
\midrule
(a). 1-CSPNet & \{$0 \sim 4$\} \\
(b). 3-CSPNet & \{$0 \sim 2$, $1 \sim 3$, $2 \sim 4$\}  \\
(c). 5-CSPNet & \{$0.0 \sim 2.0$, $0.5 \sim 2.5$, $1.0 \sim 3.0$, $1.5 \sim 3.5$, $2.0 \sim 4.0$\}\\
(d). 7-CSPNet & \{$0.0 \sim 1.0$, $0.5 \sim 1.5$, $1.0 \sim 2.0$, $1.5 \sim 2.5$, $2.0 \sim 3.0$, $2.5 \sim 3.5$, $3.0 \sim 4.0$\}\\
\bottomrule
\end{tabular}
\label{tab:configuration_time_window}
\end{table*}

\subsection{Evaluation Baselines}
The proposed approach is compared with the following diverse baselines the CSP approach FBCSP, the Riemannian-based approaches (MDM/TSM), and the DL approaches (ConvNet/EEGNet/FBCNet/SPDNet). 

\begin{enumerate}
\item FBCSP: FBCSP employs CSP on each sub-bands of EEG signals to acquire sub-band scores and then deploy the classification algorithms on selected features. 
FBCSP attained the best result in BCIC-IV-2a in 2008 and is the most representative among the CSP variants. 
The repository of the Python toolbox refers to publicly available \emph{FBCSP Toolbox} https://fbcsptoolbox.github.io/.

\item Riemannian-based Approaches: 
Minimum Distance to Riemannian Mean (MDM) and Tangent Space Mapping (TSM)~\cite{barachant2011multiclass} utilize the geodesic distances and distances of projected SPD matrices on tangent space of on \big($\mathcal{S}^n_{++}$, AIRM\big) for the EEG classification, respectively. 
For multiclass classification, we modify it using the one-versus-rest (OVR) strategy.  
The repository of the Python toolbox refers to publicly available \emph{pyRiemann} https://github.com/pyRiemann/pyRiemann.

\item Deep Learning Approaches: 
Apart from SPDNet, several CNN architectures are selected as baselines. 
ConvNet~\cite{schirrmeister2017deep} is the first CNN approach to extract the temporospatial patterns from EEG signals whose architecture consists of convolution-max-pooling blocks with a unique first convolutional layer for temporal information, standard convolution-max-pooling blocks, and a dense softmax classification layer; 
EEGNet~\cite{lawhern2018eegnet} was published soon after ConvNet, which modified CNNs concerning the properties of EEG signals, consisting of the DepthwiseConv2D layer and the SeparableConv2D layer.
Influenced by FBCSP, FBCNet~\cite{mane2021fbcnet} uses modified CNNs on each sub-bands of EEG signals to capture the temporospatiofrequency features and achieves state-of-the-art performance on several primary MI-EEG datasets;
The repository of the Python toolbox refers to the publicly available package on GitHub https://github.com/ravikiran-mane/FBCNet.

\end{enumerate}

\subsection{Naming Conventation for Hyper-parameters}
To further analyze Tensor-CSP, we introduce notations of its hyper-parameters.
Formally, $w$-CSPNet$^{(m, n, l)}$ represents the Tensor-CSPNet with $w$ window slices, $m$ banks for filters, $n$ blocks of the CSP layer, and $l$-layer neural networks in the perception layer. 
The number of banks $m$ is set to $F$, and the specific frequency ranges refer to Section~\ref{TDD}. 
The depth of the fully-connected layer $l$ has two options $\{1, 3\}$.
The output dimension of the depthwise BiMap is denoted as $o$, where $o \in$ \{4, 8, 12, 16, 20, 22, 24, 28, 32, 36\}.
The hyper-parameters in the temporal convolutional stage are a portfolio $@(p,q,r)$, where the triple is the width, the height, and the number of channels of 2D CNNs, respectively. 
The summary of the notations refers to Table~\ref{tab:Notations}.

\subsection{Evaluation Scenarios}
We will evaluate Tensor-CSPNet on two scenarios of the subject-specific analysis. 
The subject-specific analysis refers to the training and testing datasets from the same subject.
\begin{enumerate}
\item Cross-Validation Scenario (Stationary Scenario): 
This scenario uses a standard evaluation setting of 10-fold cross-validation (with a shuffle data index) for each subject. 
\item Holdout Scenario (Non-stationary Scenario):
The holdout scenario is a cross-session scenario in which the model is trained in one session and evaluated in another session. 
Figure~\ref{fig:exp_settings} illustrates the holdout scenario on two datasets. 
Note that the two-session signals of each dataset are collected on different days. 
Hence, there is typically a drift of statistical distributions between two sessions (i.e. the non-stationary phenomenon), as illustrated in Figure~\ref{fig:sub28_5_10_20_seg} {\color{blue}{(a)}}. 
\begin{figure}[!h]
\centering
\includegraphics[width=\linewidth]{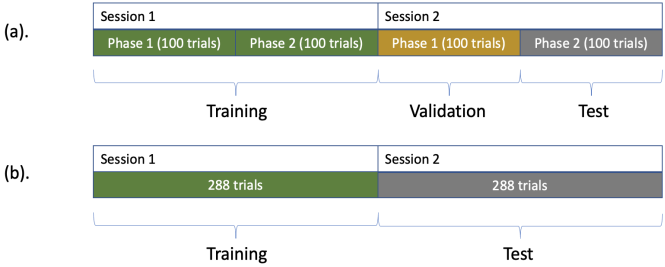}
\caption{Illustrations for Experimental Settings of the holdout scenarios on two datasets:  (a). MI-KU; (b). BCIC-IV-2a.
\label{fig:exp_settings}
} 
\end{figure}

\end{enumerate}

\begin{table*}[!t]
\caption{Average accuracies and standard deviations for the subject-specific analysis of \textbf{MI-KU} (a total of 54 Subjects) and \textbf{BCIC-IV-2a} (a total of 9 Subjects). 
Each result in the table is denoted as \emph{average accuracy (standard deviation)}.
The best-performing number for each analysis is highlighted in boldface.
}
\centering 
\begin{tabular}{l  l  l  l | l l l}
\toprule
{} & \multicolumn{3}{c|}{MI-KU (20 channels, 2 classes)} & \multicolumn{3}{c}{BCIC-IV-2a (22 channels, 4 classes)}\\
\midrule
{} & CV (S1) \% & CV (S2) \%&Holdout (S1$\rightarrow$ S2) \%& CV (T) \%& CV (E) \%& Holdout (T $\rightarrow$ E) \% \\
\midrule
FBCSP &64.41 (16.28) &66.47 (16.53) & 59.67 (14.32) & 73.57 (15.13) & 72.46 (16.02) & 65.79 (14.21)\\
MDM    &50.47 (8.63)& 51.93 (9.79)& 52.33 (6.74) & 62.96 (14.01) &59.49 (16.63) & 50.74 (13.80)\\
TSM     &54.59 (8.94)& 54.97 (9.93)&  51.65 (6.11) & 68.71 (14.32) &63.32 (12.68) & 49.72 (12.39)\\
\hline
SPDNet  &57.88 (8.68)  &58.88 (8.68) & 60.41 (12.13) &65.91 (10.31)& 61.16 (10.50)& 55.67 (9.54)\\ 
EEGNet   &63.35 (13.20) & 64.86 (13.05)& $63.28$ (11.56) & 69.26 (11.59) & 66.93 (11.31)& 60.31 (10.52)\\
ConvNet &64.21 (12.61) & 62.84 (11.74) & $61.47$ (11.22) & 70.42 (10.43) &65.89 (12.13)& 57.61 (11.09)\\
FBCNet   &74.16 (12.60) & 73.81(13.99) & $67.83$ (14.34)  & \textbf{77.26} (14.82) & \textbf{76.58} (13.09) &  72.71 (14.67) \\
Tensor-CSPNet & \textbf{74.95} (15.27)  & \textbf{75.92} (14.63) & \textbf{69.65} (14.97)  & 75.98 (14.26) & 74.92 (14.63) & \textbf{72.96} (14.98)\\
\bottomrule
\end{tabular}
\label{tab:tab_acc}
\end{table*}

\subsection{Performance Comparison}
In this section, we evaluate Tensor-CSPNet on MI-KU and BCIC-IV-2a.
Each dataset has three scenarios, including two 10-fold-cross-validation (CV) scenarios and one holdout scenario. 

The configurations for Tensor-CSPNet are a little different in each scenario. 
We require the output dimension of the depthwise BiMap layer to be $o=20$ and 22, respectively, on two datasets. 
The reason for picking such a hyper-parameter is discussed in Appdenix~\ref{hyper_parameters}. 
For the CV scenarios of both datasets, Tensor-CSPNet adopts a shallow neural network 5-CSPNet$^{(9, 1, 1)}$ because the amount of trials for training is small (i.e., 90 trial/class on MI-KU and 65 trial/class on BCIC-IV-2a), which yields the over-fitting for an extensive neural network.
For the holdout scenario of both datasets, we also adopt shallow neural networks but with finner temporal segmentation 10-CSPNet$^{(9, 1, 1)}@(9, 5, 2)$ and 5-CSPNet$^{(9, 1, 1)}@(9, 5, 4)$, respectively.
Because finner temporal segmentation is much helpful to the performance against the nonstationarity, cautiously discussed in Section~\ref{Dis:nonstationarity}.

In EEG-based BCIs, the performance of a classifier typically varies widely in different data preparation, such as the segment length of signals, number of electrodes, and electrode placements, even in the same experimental scenario. 
FBCSP is always regarded as the most stable and convincing baseline in most cases. 
From Table~\ref{tab:tab_acc}, we notice that the Riemannian-based approaches, MDM and TSM, perform like a random guess on MI-KU but a bit better on BCIC-IV-2a. 
It exhibits the limited effectiveness of using geometric quantities on SPD manifolds as the high-level features for classification. 

The mainstream DL methodology in the MI-EEG classification exploits EEG signals' temporospatiofrequency features. 
Hence, we will categorize the five DL approaches in Table~\ref{tab:tab_acc} into three groups, 
\begin{enumerate}
\item SPDNet: It only exploits the spatial patterns of EEG signals and achieves the worst performance among all the DL approaches in Table~\ref{tab:tab_acc}.
\item EEGNet and ConvNet: They exploit the temporospatial patterns, and their performances are close to FBCSP.
Note that FBCSP extracts the temporospatiofrequency patterns. 
The similar performance shows that combining any two components nearly contributes to the classification.
\item FBCNet and Tensor-CSPNet: They exploit the temporospatiofrequency patterns and outperform EEGNet and ConvNet in all scenarios, attributed to bandpass filters that embody the frequency information.
Tensor-CSPNet performs slightly better than FBCNet on MI-KU but somewhat worse on BCIC-IV-2a, except for its holdout scenario.
We briefly discuss why it performs well on both holdout scenarios in Section~\ref{Dis:nonstationarity}.
\end{enumerate}


\begin{figure*}[!t]
\centering
\includegraphics[width=0.8\textwidth]{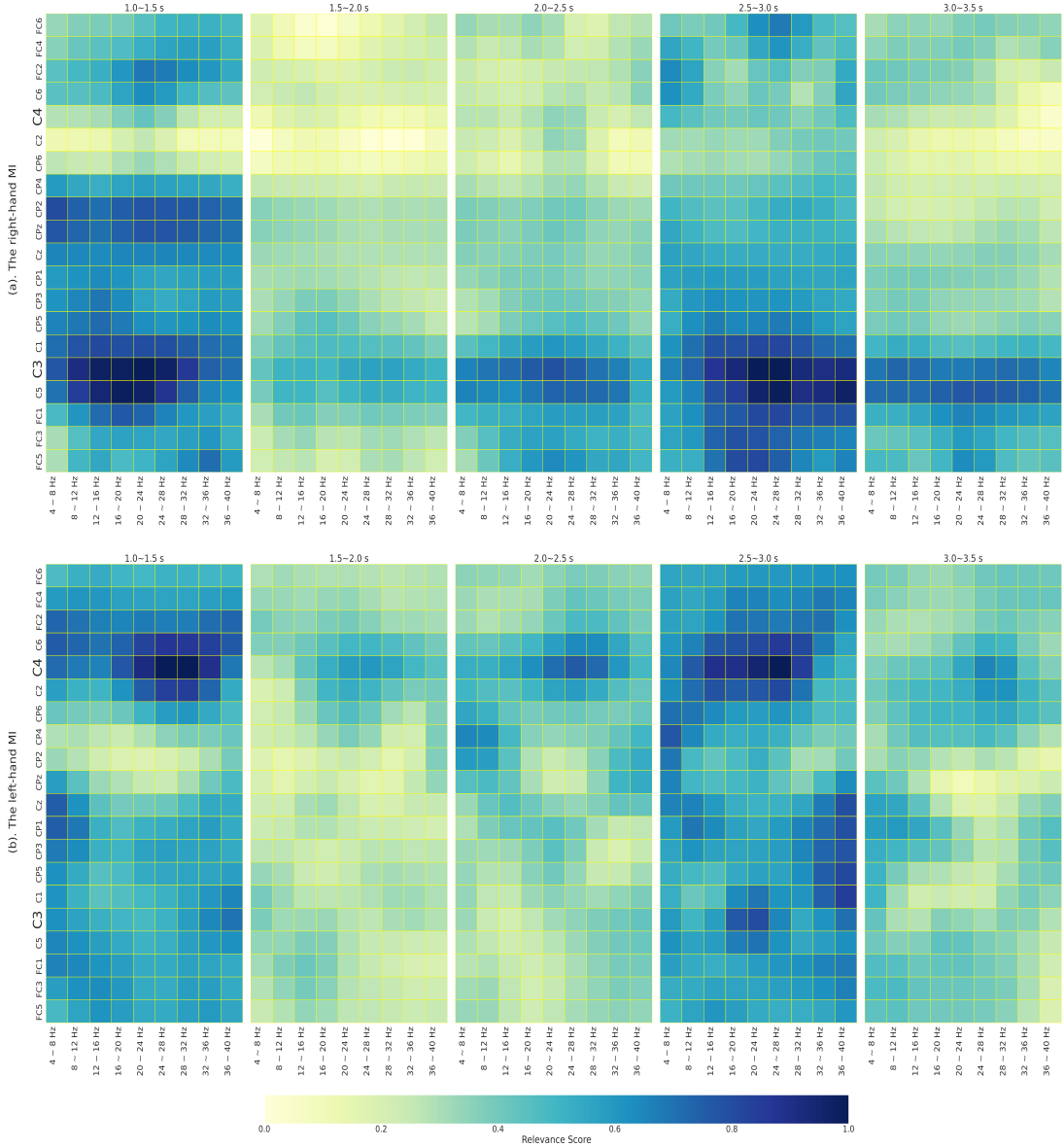}
\caption{Illustration of the heatmap of the relevance patterns of $5$-CSPNet$^{(9, 1, 1)}$:
The experiment is conducted on Subject No.2 of MI-KU with a testing accuracy of over 0.9.
The relevance pattern after DeepLIFT has an output shape (5, 9, 20, 20).
We flatten the relevance pattern into five rectangles with a height of 20 grids (20 channels) and a width of 9 grids (9 frequency bands).
Each rectangle represents the spatial-frequency information within a time window of \{$1.0 \sim 1.5$ s,  $1.5 \sim 2.0$ s, $2.0 \sim 2.5$ s, $2.5 \sim 3$ s, $3.0 \sim 3.5$ s\}. 
The rectangle column records the main diagonal of the relevance pattern's $20 \times 20$ covariance matrix.
The value in each cell on the heatmap is normalized in [0, 1] and smoothed by a Gaussian filter.
\label{fig:relevance}
} 
\end{figure*}

\subsection{Interpretability Analysis}
In this section, we investigate the interpretability of extracted temporospatiofrequency patterns of Tensor-CSPNet using Deep Learning Important FeaTures (DeepLIFT)~\cite{shrikumar2017learning}, which is a gradient-based interpretation method widely employed in the BCI classification~\cite{lawhern2018eegnet,mane2021fbcnet}. 

To interpret the extracted features, we propose a simple visualized approach to flatten the four-dimensional relevant pattern of DeepLIFT to a two-dimensional rectangle, illustrated in Figure~\ref{fig:relevance}.
Subject No.2 is selected from the MI-KU dataset for interpretation, whose testing accuracy is over 0.9. 
The upper and lower rows in the heatmap represent the right-hand and left-hand MI, respectively, and are interpreted as follows, 
\begin{enumerate}[(a)] 
\item The right-hand MI: Patterns with 8$\sim$28 Hz highlights around C3 in 1.0$\sim$1.5 sec and 2.5$\sim$3.0 sec. 
\item The left-hand MI: Patterns with 24$\sim$28 Hz highlights around C4 in 1.0$\sim$1.5 sec and 2.5$\sim$3.0 sec.
\end{enumerate}
The above-interpreted temporospatiofrequency information is consistent with the existing practical frequency components of the left and right-hand MI~\cite{pfurtscheller1997eeg} that the alpha band 9$\sim$14 Hz and beta bands 18$\sim$26 Hz perform more significantly on C3 and C4 of the primary motor cortex, or M1. 
Two active time windows indicate the change of band power event-related desynchronization (ERD) and the event-related synchronization (ERS) occurring during MI~\cite{pfurtscheller1999event}.

\begin{figure*}[!t]
\centering
\includegraphics[width=\linewidth]{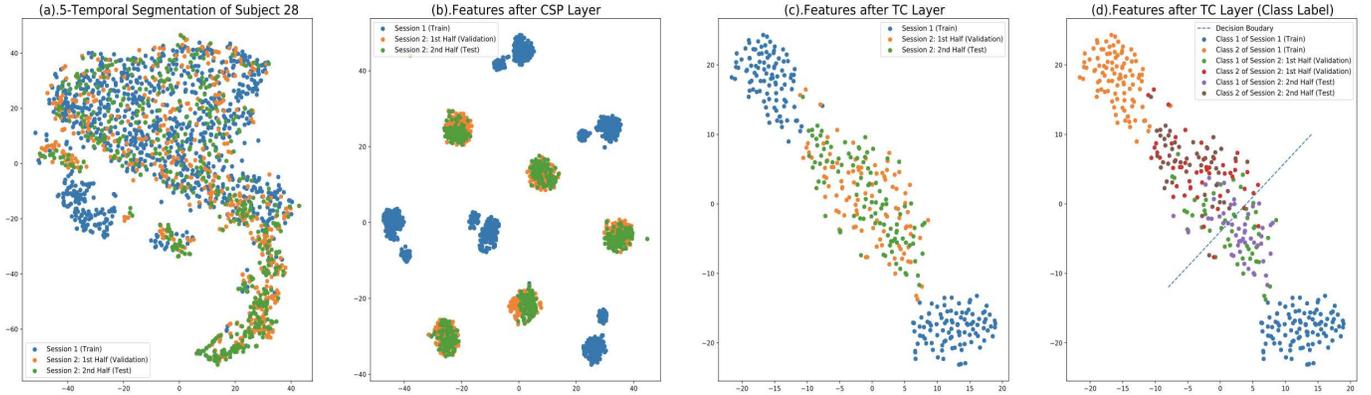}
\caption{Illustration of outputs of each intermediate stage in 5-CSPNet$^{(9, 1, 1)}$ with $o=22$ on Subject No.28 of MI-KU:
The time windows for the model are $1 \sim 1.5$ s, $1.5 \sim 2.0$ s, $2.0 \sim 2.5$ s, $2.5 \sim 3.0$ s, and $3.0 \sim 3.5$ s.
(a). 5-Temporal Segmentation of Subject No.28: This figure is the same as Fig.~\ref{fig:sub28_5_10_20_seg}, but there is a rotation due to the figure scale. 
(b). Features after CSP layer (Stage 2): Blue and yellow/green have five segments because of the five temporal segmentation. We name each data cluster as the temporal segment in this paper. 
(c). Features after TC layer (Stage 3): Segments of either blue or yellow/green aggregates. The yellow/green one lies in the middle of two blue parts. 
(d). Features after TC layer (Class Label): Draw the points with label information. Two classes are almost evenly distributed on both sides of the decision boundary.
\label{fig:tensor_cspnet_4fig}
} 
\end{figure*}

\subsection{Visualization}
In this section, we plot the 2-dimensional projections for outputs of each intermediate layer in Tensor-CSPNet using t-distributed Stochastic Neighbor Embedding (t-SNE)~\cite{van2008visualizing}.
The t-SNE algorithm is a widely used technique of non-linear dimensionality reduction to visualize high-dimensional data.
Specifically, we will investigate the mechanism of Tensor-CSPNet via visualizing the outputs of each intermediate layer in the holdout scenario of MI-KU.
Subject No.28 in MI-KU is chosen for visualizing because Tensor-CSP attains a good performance on this subject with an accuracy of 0.92.
Attributed to the tensor stacking stage, the training set is mixed up with the validation and test sets, as illustrated in Figure~\ref{fig:tensor_cspnet_4fig} {\color{blue}(a)}.
The CSP stage centralizes the data shape in each temporal segment of the training, validation, and test sets, respectively, so that there are fifteen segments (= 5 $\times$ the training/validation/test sets.) in Figure~\ref{fig:tensor_cspnet_4fig} {\color{blue}(b)}.
The temporal concentration and 2D CNN in the TC stage concentrate temporal segments of either the training, validation, or test sets along the time domain, as illustrated in Figure~\ref{fig:tensor_cspnet_4fig} {\color{blue}(c)}.
The Figure~\ref{fig:tensor_cspnet_4fig} {\color{blue}(d)} records the label-wise projections in which we can distinguish the different class of labeled data.
In addition, we notice that data with Class 1 and Class 2 lies on the bottom and upper sides of the decision boundary, respectively. 
More examples and visualization of BCIC-IV-2a are illustrated in Appendix.

\section{Discussion}\label{Discussion}
In the discussion, we first provide evidence of why Tensor-CSPNet outperforms the other approaches in the non-stationary scenarios. 
Then, we will discuss the relationship between Tensor-CSPNet and other existing mainstreams of the MI-EEG classifiers.
\subsection{Evidence of Temporal Segmentation against Non-stationarity}\label{Dis:nonstationarity}
\begin{figure}[!h]
\centering
\includegraphics[width=\linewidth]{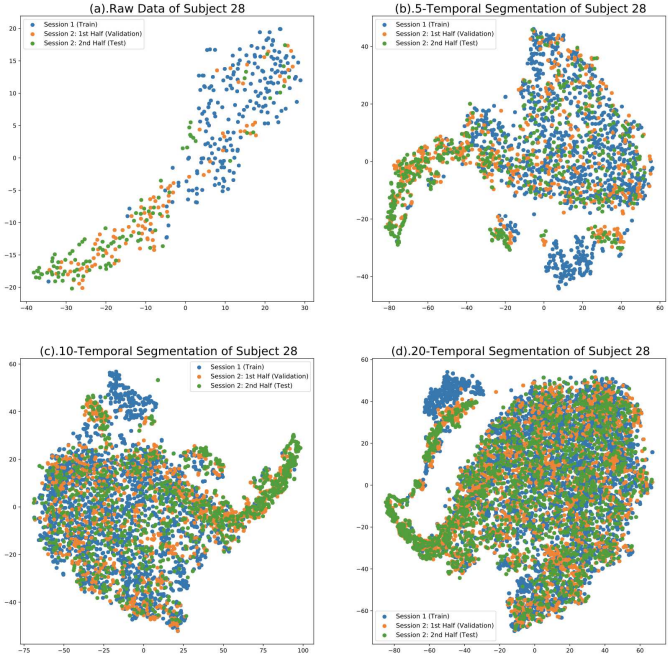}
\caption{2-dimensional Projection of Subject No.28 in MI-KU using t-SNE:
There are two sessions for each subject in the MI-KU dataset. 
S1 is for the training set, and two halves of S2 are for the validation and test sets. 
The lengths of time windows are (a).2500 ms, (b).500 ms, (c).250 ms, and (d).125 ms.  
There is no overlapping between time windows. 
Each 2-dimensional color point is dimensionality reduced from a $9 \times 20 \times 20$-dimensional point, where it has 20 electrodes in the motor cortex region and nine frequency bands. 
This is the input format for \emph{Tensor-CSPNet}.
\label{fig:sub28_5_10_20_seg}
} 
\end{figure}

Early electrophysiological studies show that large-scale patterns of synchronized neuronal activity exhibit considerable variability over time, e.g., alpha-blocking with eyes opening, the transition from wakefulness to drowsiness, etc.
The variability was termed as the nonstationarity nature of EEG signals~\cite{kaplan2005nonstationary} and mainly caused the drift in statistical distribution between different sessions and subjects.
To determine Tensor-CSPNet's good performance in non-stationary scenarios, we pick Subject No.28 of MI-KU because Tensor-CSPNet's accuracy of this subject is 0.3 higher than FBCSP's. 
Figure~\ref{fig:sub28_5_10_20_seg} exhibits a noticeable trend that the more refined temporal segmentation yields a more extensive crossover region of the training, validation, and test sets in the statistical distribution space. 
In the view of statistics, temporal segmentation fixes the nonstationarity, which is the rediscovery of a four-decade-ago theory called \emph{segmentation techniques} for nonstationary EEGs~\cite{barlow1985methods}.
In the view of neural signals, the fixed-interval temporal segmentation breaks down EEG signals into many short piecewise quasi-stationary intervals. Therefore, the drift between different sessions disappears in the numerical aspect, which is helpful to classification performance when using the statistical classifier.

\subsection{Tensor-CSPNet VS. Other BCI Classifiers}

\subsubsection{DL}
Most of the DL approaches in the MI-EEG classification are designed to exploit the temporospatial information from EEG signals using CNNs. 
In contrast, Tensor-CSPNet formulates EEG signals on SPD manifolds, uses existing layers in SPDNet on SPD manifolds to exploit the spatial patterns from SCMs, and uses CNNs to capture the temporal dynamics of EEG signals on the tangent space. 
\subsubsection{CSP}
Attributed to the BiMap layer, Tensor-CSP performs like a CSP-like approach.
The weight update using the data-driven approach improves the knowledge that the most \emph{appropriate} projection matrix $W$ can be entirely determined by label data rather than using the rule of simultaneous diagonalization. 
Moreover, Riemannian BN performs a regularization in Tensor-CSPNet similar to the Regularized-CSP approach~\cite{lotte2010regularizing}, and ReEig leverages the linear spatial filter to a non-linear one. 
\subsubsection{Riemannian-Based Approaches}
Both Tensor-CSPNet and the Riemannian-based approach characterize EEG signals on SPD manifolds. 
The Riemannian-based approach uses geodesic distance on SPD manifolds as a high-level feature for the MI-EEG classification. 
In contrast, Tensor-CSPNet uses the low-level feature expressions of SMCs captured by a neural network-based approach for classification. 
\subsubsection{Manifold Learning}
Manifold learning~\cite{belkin2001laplacian} is a theoretical dimensionality reduction setting in which the samples are assumed to be on or near a low-dimensional submanifold embedding in high-dimensional space. 
It aims to acquire a low-dimensional geometric representation of high-dimensional data retaining a meaningful property.
Architecture SPDNet can be regarded as a new class of manifold learning for the supervised learning setting because it is a neural-network-based transformation from one SPD manifold to another, and so is Tensor-CSPNet. 
However, the studies in Appendix~\ref{outputdimension} exhibit that expanding the dimension, rather than reducing it, yields a better classification performance in some cases. 

\section{Conclusions}
In this work, we propose a novel GDL framework called Tensor-CSPNet to exploit the temporospatiofrequency features of EEG signals for a general EEG-BCI classification paradigm. 
To achieve this goal, the framework is inspired by a growing interest in formulating EEG signals on SPD manifolds and uses existing network architectures on SPD manifolds to exploit the patterns.
Tensor-CSPNet exhibits better classification performance in the experiments than the current state-of-the-art approach.
In addition, we investigate how each layer in Tensor-CSPNet works and how temporal segmentation improves the Tensor-CSPNet's performance in the cross-session scenario and gives an interpretability analysis of the extracted patterns. 
The current experimental results demonstrate the validity of Tensor-CSPNet for the MI-EEG classification. 
Despite the validity, Tensor-CSPNet also has the following appealing upsides to existing CNN classifiers: 
1). SPD-matrix representation for encoding spatial patterns is typically compact and robust to noise.
2). Specific Architecture on SPD manifolds to enhance feature extraction. 
For example, It preserves the SPD structure of matrices across layers and essentially maintains more encoding information of SCM.
In addition, the combination of the depthwise BiMap layer and ReEig improve the feature expression of spatial patterns. 
3). Tensor stacking for well-performing against nonstationarity.

\section{Acknowledgment}
This study is supported under the RIE2020 Industry Alignment Fund–Industry Collaboration Projects (IAF-ICP) Funding Initiative, as well as cash and in-kind contributions from the industry partner(s). 
This study is also supported by the RIE2020 AME Programmatic Fund, Singapore (No. A20G8b0102).

{\footnotesize
\bibliographystyle{IEEEtran}
\bibliography{refs}
}

\appendices

\section{Prior Assumptions for CNNs}~\label{assumption}
In this section, we briefly introduce the prior assumptions on the data domain with which the CNN-type architecture can effectively extract the local statistics from data. 
For a more in-depth review of these assumptions, we refer the readers to~\cite{bruna2013invariant,bronstein2017geometric} and references therein. 
Technically, suppose a signal $\phi(x) \in \mathcal{L}^2(\Omega)$, where $x \in \Omega \subset \mathbb{R}^d$.
The goal of the supervised learning setting is to train a statistical model $f:\mathcal{X} \mapsto \mathcal{Y}$, where $\mathcal{X}$ is the space of representations $\phi(x)$ and $\mathcal{Y}$ is typically a discrete set of labels.
We say model $f$ is \emph{translation invariant} and \emph{translation equivariant} with respect to any $\phi \in \mathcal{L}^2(\Omega)$ and any $v\in\Omega$ if $f\big( \phi(x-v)\big) = f \big(\phi(x)\big)$ and $f\big( \phi(x-v)\big) = f \big(\phi(x)-v\big)$ respectively. 
Many tasks in computer vision are assumed to be \emph{translation invariant} and \emph{translation equivariant} and required to be \emph{stable} with respect to local deformations that is defined as a Lipschitz continuity condition as follows, 
\[
||f\big( \phi(x - \tau(x)) \big)  - f \big(\phi(x)\big)|| \leq C \cdot ||\phi||_{\mathcal{L}^2} \cdot \sup |\nabla \tau(x)|,
\]
where $C$ is constant, $\tau(x)$ is a smooth displacement field that deforms the signal, and $\nabla \tau(x)$ is the deformation gradient tensor.

\section{SPD Mainfolds}~\label{appen:metric}
In this section, we give a brief overview of SPD manifolds with respect to the affine invariant Riemannian metric (AIRM)~\cite{skovgaard1984riemannian,pennec2006riemannian} and the weighted Riemannian barycenter. 
For a more in-depth review of the geometry of the space of $\mathcal{S}^n_{++}$, we refer the reader to \cite{petersen2006riemannian,bhatia2009positive} and references therein.

\subsection{Riemannian Geometry of SPD Matrices}
The space of $ \mathcal{S}^n_{++}$ is a Riemannian manifold if endowed with a Riemannian metric.
AIRM, a widely-used class of the Riemannian metric for the space of $ \mathcal{S}^n_{++}$, was put forward independently from information science in the 1980s~\cite{burbea1982entropy} and engineering disciplines~\cite{pennec2006riemannian,fletcher2007riemannian} after 2005.
Formally, AIRM is defined as $g_{P}(v, w):= \langle \, P^{-\frac{1}{2}} v P^{-\frac{1}{2}}, P^{-\frac{1}{2}} w P^{-\frac{1}{2}} \, \rangle_{\mathcal{F}}$, for each $v$ and $w$ on tangent space $\mathcal{T}_{P} \mathcal{S}^n_{++}$. 
Riemannian manifold \big($\mathcal{S}^n_{++}$, AIRM\big) is a Hadamard that is simply connected and complete with everywhere non-positive sectional curvature.
It holds many nice properties, for example, there is an unique \emph{geodesic}\footnote{\, The geodesic on Riemannian manifolds $(M, g)$ with respect to the Levi-Civita connection $\nabla$ is defined as a curve $\gamma (t)$ such that $\nabla_{\dot{\gamma}} \dot{\gamma} = 0$.}
$\gamma (t): [0, 1] \longmapsto \mathcal{S}^n_{++}$ between any two SPD matrices $P_1$ and $P_2$ of $\mathcal{S}^n_{++}$ such that $\gamma (0) := P_1$, $\gamma (1) := P_2$ and 
$
\gamma (t) := P_1^{\frac{1}{2}} \cdot (P_1^{-\frac{1}{2}} \cdot P_2 \cdot P_1^{-\frac{1}{2}})^t \cdot P_1^{\frac{1}{2}}
$
with the arc-length $\mathcal{L}_g (\gamma) = ||\log(P_1^{-1/2} \cdot P_2 \cdot P_1^{-1/2})||_{\mathcal{F}}$.
In addition, the geodesic distance on \big($\mathcal{S}^n_{++}$, AIRM\big) is invariant under any congruence transformations $\Gamma_W$, i.e., $\mathcal{L}_g \big( \Gamma_W \circ \gamma \big) = \mathcal{L}_g (\gamma)$.
A parallel transport on \big($\mathcal{S}^n_{++}$, AIRM\big) $\Gamma_{P_1 \rightarrow P_2}:\mathcal{T}_{P_1} \mathcal{S}^n_{++} \longmapsto \mathcal{T}_{P_2} \mathcal{S}^n_{++}$ is given by $\Gamma_{P_1 \rightarrow P_2}(v) := (P_2 P_1^{-1})^{\frac{1}{2}} v (P_2 P_1^{-1})^{\frac{1}{2}}$, where $P_1, P_2 \in \mathcal{S}^n_{++}$ and $v \in \mathcal{T}_{P_1} \mathcal{S}^n_{++}$.

\subsection{Weighted Riemannian Barycenter}
The weighted Riemannian barycenter is a generalization of weighted barycenter on Riemannian manifolds. 
In our study, the Riemannian-based BCI classifier and Riemannian BN have been used in the computation procedure. 
Formally, given a batch $\mathcal{B}$ of $N$ SPD matrices $\{P_i\}_{i=1}^N$, the weighted Riemannian barycenter (a.k.a. $Fr\acute{e}chet$ mean~\cite{frechet1948elements}) on \big($\mathcal{S}^n_{++}$, AIRM\big) is given as the solution to the following optimization problem~\cite{bacak2014computing}:
\[
Bar_w (\mathcal{B}) := \arg\min_{M \in \mathcal{S}^n_{++}} \sum_{i=1}^N w_i \cdot \mathcal{L}_g (M, P_i)^2, 
\]
where weights $w_i \geq 0$ $(i = 1, ..., N)$ and $\sum_{1\leq i \leq N} w_i = 1$.
%

\section{Fixed-Interval Segmentation and q value}~\label{appen:fixed} 
In this section, we propose a strategy for those signals that we are not familiar with their temporal characteristics. 
We call this strategy to be fixed-interval segmentation.
Technically, this strategy means that the EEG signals are initially subdivided into fixed, short equal-length intervals or segments without overlapping.
In the main paragraph, this strategy is used for the evaluation of the MI-KU dataset in which we require that the length of each time window can divide by the length of EEG signals for simplicity, and therefore, we employ two configurations, 5-CSPNet, and 10-CSPNet, whose length of time windows is 500 ms and 250 ms, respectively on MI-KU.
Furthermore, we conclude that the best q value under the strategy of the fixed-interval segmentation is equal to width W. (q is the height of 2D CNN in the TC stage.)
This is because the model performance monotonically increases as the $q$ value increases in Table~\ref{tab:strategy}.
It is also consistent with the neurobiological fact that a wider window size yields a higher probability of examining event-related desynchronization/synchronization during motor imagery tasks.

\begin{table}[!h]
\caption{The results in the hold-out scenario of MI-KU with different $q$ values in the TC stage of 5-CSPNet and 10-CSPNet.
q is the height of 2D CNN in the TC stage. 
\label{tab:strategy}
}
\centering 
\begin{tabular}{l  r  r  r  r  r}
\toprule
5-CSPNet &  q=1  & q=2 & q=3 & q=4 & q=5\\
\midrule
Acc. & 0.635 & 0.651 & 0.660 & 0.668 & \textbf{0.676}\\
\midrule
10-CSPNet &  q=6  & q=7 & q=8 & q=9 & q=10\\
\midrule
Acc. & 0.668 & 0.674 & 0.678 & 0.672 & \textbf{0.684}\\
\bottomrule
\end{tabular}
\end{table}

\begin{table*}[!t]
\caption{
Notations for Hyper-Parameters in $w$-CSPNet$^{(m, n, l)}$ $@(p,q,r)$. 
The different configurations of temporal segments on two datasets refer to Table~\ref{tab:configuration_time_window}.
}
\centering 
\begin{tabular}{l  l l }
\toprule
Hyper-Parameters   & Meaning   &Portfolio of Pre-set Parameters \\
\midrule
$w$  & The number of time window slices.  & $w \in \{1, 5, 10\}$ in MI-KU, and $w \in \{1, 3, 5, 7\}$ in BCIC-IV-2a. \\
\hline
$m$  & The number of bandpass filters. & $m = 9$. \\
$n$   & The number of the CSP stages. & $n \in \{1, 3\}$.  \\
$l$    & Depth of the fully-connected networks in the classification stage. & $l \in \{1, 3\}$.\\
\hline
$o$ & The output dimension $o$ of the CSP stage.  &   $o\in \{4, 8, 12, 16, 20, 22, 24, 28, 32, 36\}$.\\
$(p, q, r)$ & The width, height, and output channels in 2D CNN of the TC stage. & $p \in \{1, 9\}$.\\
\bottomrule
\end{tabular}
\label{tab:Notations}
\end{table*}

\section{Ablation Study on BCIC-IV-2a}\label{hyper_parameters}
This section investigates the effects of each layer and hyper-parameter of Tensor-CSPNet on the training session of BCIC-IV-2a.
We will have an in-depth analysis of its mechanism using visualization and discuss the computational efficiency of the Tensor-CSPNet. 
For ease of communication, we summarize the naming conventions for hyper-parameters in Tensor-CSPNet in Table~\ref{tab:Notations}.
Primarily, we put a symbol BN at the end of each configuration of Tensor-CSPNet to indicate if it has Riemannian BN in the CSP stage. 

\subsection{Output Dimension $o$ of the Depthwise BiMap Layer}\label{outputdimension}
\begin{figure}[!h]
\centering
\includegraphics[width=9cm]{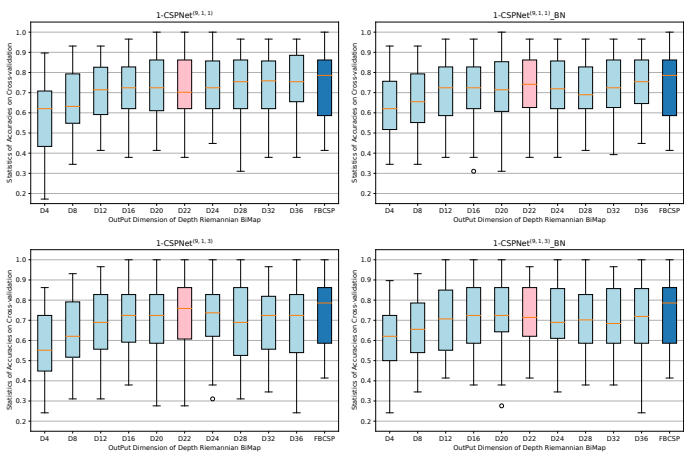}
\caption{Box Plots for Output Dimension of the depthwise BiMap:
Box plots of the statistics of outputs for 1-CSPNet$^{(9, 1, 1)}$,  1-CSPNet$^{(9, 1, 1)}$\_{BN}, 1-CSPNet$^{(9, 1, 3)}$ and 1-CSPNet$^{(9, 1, 3)}$\_{BN} for various output dimensions \{4, 8, 12, 16, 20, 22, 24, 28, 32, 36\} of the depthwise BiMap on BCIC-IV-2a. 
Baseline (FBCSP) is in dark blue and 22 ( = channel dimension) is in pink. 
\label{box_1}
} 
\end{figure}

We investigate the output dimension $o$ in the depthwise BiMap layer. 
The average accuracies and standard deviations for evaluation are summarized in Table~\ref{tab:acc_table}, and their quartiles are box-plotted in Figure~\ref{box_1}.
Based on the table and quartile, we have the following observations: 
\begin{enumerate}
\item Observation One (O1): 
The average accuracy monotonically increases as the output dimension $o$ increases.
In Table~\ref{tab:acc_table}, the Dimension of 36 has statistically significantly better accuracies than other output dimensions across four architectures (p<0.05, Wilcoxon signed-rank test).
When the output dimension is over half the channel dimension, the average accuracy is not statistically significantly different from the FBCSPs across four architectures (p>0.3, Wilcoxon signed-rank test). 
But, it is statistically significantly lower than FBCSP when the output dimension is less than half across four architectures (p<0.01, Wilcoxon signed-rank test). 
\item Observation Two (O2): The expansion yields slightly better average accuracy than FBCSP, which might improve a specific subject.
Dimension of 32 and 36 of both 1-CSPNet$^{(9, 1, 1)}$ and 1-CSPNet$^{(9, 1, 1)}$\_{BN} have an improvement on average accuracies by almost $1\sim 2\%$ than FBCSP.
Particularly, Dimension of 36 of 1-CSPNet$^{(9, 1, 1)}$ achieves 0.64 for Subject No.2, whereas FBCSP has 0.52 (p=0.074, Wilcoxon signed-rank test).
\item Observation Three (O3): The architecture with a multilayer statistically significantly performs worse than the single layer when the output dimension o is over 22 (average p<0.05, Wilcoxon signed-rank test).
\end{enumerate}

\begin{table*}[!h]
\caption{Experiments on the Output Dimension $o$ of the Depthwise BiMap Layer: Average accuracy (Acc.) and standard deviation (Std.) under 1-CSPNet$^{(9, 1, 1)}$, 1-CSPNet$^{(9, 1, 1)}$\_{BN}, 1-CSPNet$^{(9, 1, 3)}$, and 1-CSPNet$^{(9, 1, 3)}$\_{BN} for a variety of output dimensions in the depthwise BiMap layer on BCIC-IV-2a. 
In this case, the accuracy and standard of FBCSP are 0.7357 and 0.1513, respectively. 
The best-performing method for each analysis is highlighted in boldface.
}
\centering 
\begin{tabular}{l l r r r r r |r| r r r r |r}
\toprule
Algorithm/Dimension &  Metric & 4 &      8 &     12 &     16 &     20 &     22 &     24 &     28 &     32 &   36  & Avg.\\
\midrule
1-CSPNet$^{(9, 1, 1)}$ &Acc. & 0.5793 &0.6559&0.7023&0.7246&0.7316&0.7304&0.7385&0.7326&0.7408&0.7496 & 0.7086\\
&Std. & 0.1572&0.1296 &0.1218&0.1225&0.1387&0.1306&0.1303&0.1365&0.1343&0.1294 & 0.1331\\
\midrule
1-CSPNet$^{(9, 1, 1)}$\_{BN} &Acc. & 0.6400 &0.6671&0.7041&0.7214&0.7172&0.7383&0.7331&0.7096&0.7413&0.7508 & \textbf{0.7123}\\
&Std. & 0.1241&0.1324 &0.1331&0.1288&0.1430&0.1276&0.1284&0.1231&0.1321&0.1329 & \textbf{0.1305}\\
\midrule
1-CSPNet$^{(9, 1, 3)}$ &Acc. & 0.5739 &0.6351&0.6814&0.7159&0.7034&0.7181&0.7191&0.6889&0.6927&0.6823 & 0.6811\\
&Std. & 0.1339&0.1405 &0.1469&0.1448&0.1559&0.1443&0.1369&0.1577&0.1496&0.1603 & 0.1471\\
\midrule
1-CSPNet$^{(9, 1, 3)}$\_{BN}  & Acc. & 0.6116 &0.6493&0.6975&0.7134&0.7287&0.7277&0.7124&0.7011&0.7023&0.6972 & 0.6941\\
& Std. & 0.1205&0.1371 &0.1444&0.1454&0.1356&0.1258&0.1387&0.1429&0.1385&0.1527 &0.1381\\
\bottomrule
\end{tabular}
\label{tab:acc_table}
\end{table*}

\begin{table*}[!h]
\caption{Experiments on Effectiveness of Riemannian BN: 
Average accuracy (Acc.) and standard deviation (Std.) under four pairs of CSPNets with the CSPNet$^{(9, 1, 1)}$ or the CSPNet$^{(9, 3, 1)}$ on BCIC-IV-2a. 
The best-performing method for each analysis is highlighted in boldface.
}
\centering 
\begin{tabular}{l  r | r | r | r}
\toprule
{} &  1-CSPNet$^{(9, 1, 1)}$  & 1-CSPNet$^{(9, 1, 1)}$ \_{BN}   & 1-CSPNet$^{(9, 1, 3)}$  &  1-CSPNet$^{(9, 1, 3)}$ \_{BN}  \\
\midrule
Acc. & \textbf{0.7304}&\textbf{0.7383}&\textbf{0.7181}&\textbf{0.7277}\\
Std. & 0.1306&0.1276&0.1443&0.1258\\
\midrule
{} &  1-CSPNet$^{(9, 3, 1)}$   &  1-CSPNet$^{(9, 3, 1)}$ \_{BN}   &  1-CSPNet$^{(9, 3, 3)}$   &  1-CSPNet$^{(9, 3, 3)}$ \_{BN}  \\
\midrule
Acc. & 0.7292&0.7327&0.6945&0.7241\\
Std. &0.1341 &0.1267&0.1408&0.1381\\
\bottomrule
\end{tabular}
\label{tab:appendix_table6}
\end{table*}

\begin{remark}
From Table~\ref{tab:acc_table}, Tensor-CSPNet reveals an interesting phenomenon: the accuracy will improve when we lift the output dimension of the depthwise BiMap. 
The map from a small input to a large output is like a volume-conduction problem reconstructing EEG sources. 
It is evident that many latent variables exist in the EEG signals because the number of current sources is significantly greater than measurements in the 3-dimensional brain volume of the electrophysiological source imaging in neurophysiology.~\cite{he2018electrophysiological}
\end{remark}

\subsection{Validity of the Riemannian BN}
The Riemannian BN is after depthwise BiMap and before ReEig inspired by the position of BN in ResNet~\cite{he2016deep}.
According to Table~\ref{tab:acc_table}, we have Observation Four (O4), that Riemannian BN relieves overfitting and improves up to 1\% on the average accuracy in many pairs.
Notice that the average accuracy and standard deviation across all the output dimensions 0.7123 and 0.1305 are both better than the ones of 1-CSPNet$^{(9, 1, 1)}$, respectively, so with the pair of 1-CSPNet$^{(9, 1, 3)}$\_{BN} and 1-CSPNet$^{(9, 1, 3)}$.
However, the improvement has no statistical significance, i.e., average p>0.4 across 20 top-bottom pairs in Table~\ref{tab:acc_table}, Wilcoxon signed-rank test.
Figure~\ref{box_1} exhibits this statistical result that the shapes of quartiles for both kinds of architecture are similar. 

\begin{table*}[!h]
\caption{Experiments on Time-Frequency Resolution:
Average accuracy (Acc.) and standard deviation (Std.) of Tensor-CSPNet under four configurations of the time windows of 1-CSPNets$^{(9, 1, 1)}$\_{BN} without the temporal convolutional stage on BCIC-IV-2a.
The best-performing method for each analysis is highlighted in boldface.
}
\centering 
\begin{tabular}{lrrrr}
\toprule
Sub./Architecture  &  1-CSPNets$^{(9, 1, 1)}$\_{BN} &  3-CSPNets$^{(9, 1, 1)}$\_{BN} & 5-CSPNets$^{(9, 1, 1)}$\_{BN} &    7-CSPNets$^{(9, 1, 1)}$\_{BN} \\
\midrule
Acc. & {\bf 0.7383}& 0.7238& 0.7334 &0.6821\\
Std. & 0.1276&0.1309 &0.1320&0.1357\\
\bottomrule
\end{tabular}
\label{tab:appendix_table8}
\end{table*}

\subsection{Depth of Architecture}
For the sake of brevity, the output dimension $o$ of each depthwise BiMaps is set 22 in a 3-block CSP layer, i.e. $o_1,o_2,o_3 = 22$. 
In Table~\ref{tab:appendix_table6}, the shallow model (top) statistically significantly outperforms the deep one (down) in each top-down pair (average p<0.05 across top-down pairs, Wilcoxon signed-rank test). 
Apart from the effect of the depth of architecture, we notice that the average accuracy of 1-CSPNet$^{(9, 1, 1)}$ and 1-CSPNet$^{(9, 3, 1)}$ (Column One) are statistically significantly better than ones of 1-CSPNet$^{(9, 1, 3)}$ and 1-CSPNet$^{(9, 3, 3)}$ (Column Three), respectively in Table~\ref{tab:acc_table} (average p<0.02 across top-down pairs, Wilcoxon signed-rank test), so with Column Two and Four (average p<0.05 across top-down teams, Wilcoxon signed-rank test).
Thus, we have Observation Five (O5), that the shallow model statistically significantly outperforms the deep one, and the single-layer statistically greatly exceeds the multilayer (average p<0.05, Wilcoxon signed-rank test).
%

\subsection{Time-Frequency Resolution}
The frequency resolution is fixed at 4 Hz as the sub-band approach for brevity. 
In particular, although period $0.0 \sim 1.0$ s after the cue is an imagination preparation stage and period $3.5 \sim 4.0$ s is a post imagination stage in BCIC-IV-2a, the first window slice begins at 0.0 s. 
In the experiments, we pick architecture 1-CSPNets$^{(9, 1, 1)}$\_{BN} with output dimension $o=22$, but the temporal convolutional stage is removed to get rid of the effects of the temporal dynamic behavior.
Instead, we concatenate the extracted features for the classification stage.
Hence, we have Observation Six (O6) that the pre-set temporal segmentation has no statistically significant improvement in the average accuracy (p>0.1 for 3-/5-CSPNet except 7-CSPNet, Wilcoxon signed-rank test), but it statistically significantly improves the performance of a specific subject.
Specifically, 3-CSPNets$^{(9, 1, 1)}$\_{BN}, 5-CSPNets$^{(9, 1, 1)}$\_{BN}, and 7-CSPNets$^{(9, 1, 1)}$\_{BN} statistically significantly improve almost 10$\%$ on Subject No.9 from Table~\ref{tab:appendix_table8} (p<0.02, Wilcoxon signed-rank test). 
In addition, the inappropriate segmentation, for instance, 7-CSPNets$^{(9, 1, 1)}$\_{BN}, has a clear drop of 5\% in the average accuracy.

\begin{table}[!h]
\caption {Experiments on Hyper-parameters in the Temporal Convolutional Stage: Average accuracy (Acc.) and standard deviation (Std.) of various hyper-parameter portfolios of the temporal convolutional stage for 5-CSPNets$^{(9, 1, 1)}$\_{BN} on BCIC-IV-2a. 
The portfolio $@(p,q,r)$ represents (Width, Height, Number of the Output Channels) in the temporal convolutional stage.
The best-performing result is highlighted in boldface.
}
\centering 
\begin{tabular}{lrrr }
\toprule
Portfolio & $@(9,2,1)$  & $@(9,2,10)$ &  $@(9,2,20)$\\
\midrule
Acc. & 0.5526 & 0.6896 & 0.7173\\
Std. & 0.1497 & 0.1295  & 0.1357\\
\midrule
{} &  $@(1,2,1)$   & $@(1,2,10)$ & $@(1,2,20)$\\
\midrule
Acc. & 0.6166&0.7308& \bf{0.7412}\\
Std. & 0.1512&0.1375 &0.1374\\
\bottomrule
\end{tabular}
\label{tab:appendix_table10}
\end{table}


%
\subsection{Hyper-parameters in the Temporal Convolutional Stage} 
In the temporal convolutional stage, there are three hyper-parameters in a portfolio $@(p, q, r)$, where width $p$ ($p$ = 1 or $F)$ and height $q$ $(1\leq q \leq W)$, and output channels $r$ for 2D CNN. 
According to the analysis in the previous subsection, we pick 5-CSPNets$^{(9, 1, 1)}$\_{BN} with output dimension $o=22$ in this experiment. 
The height $q$ is set at 2 in all the cases. 
The width $p$ has two options from $\{1, 9\}$, and the number of output channels $r$ have three options from $\{1, 10, 20\}$.
The more refined segmentation of hyper-parameters in the temporal convolutional stage yields a higher dimension of the output vector.
For instance, for a 5-CSPNets$^{(9, 1, 1)}$, the dimension of the concatenated vector after a 2D CNN with $@(p,q,r)=(1,2,20)$ is 720 \big($=9\times (5-1) \times 20$\big).
From Table~\ref{tab:appendix_table10}, we have Observation Seven (O7) that the performance statistically significantly monotonically improves as the number of output channels increases and the width of 2D CNN decreases (p<0.05, Wilcoxon signed-rank test).

\subsection{Visualization}
\begin{figure}[!h]
\centering
\includegraphics[width=\linewidth]{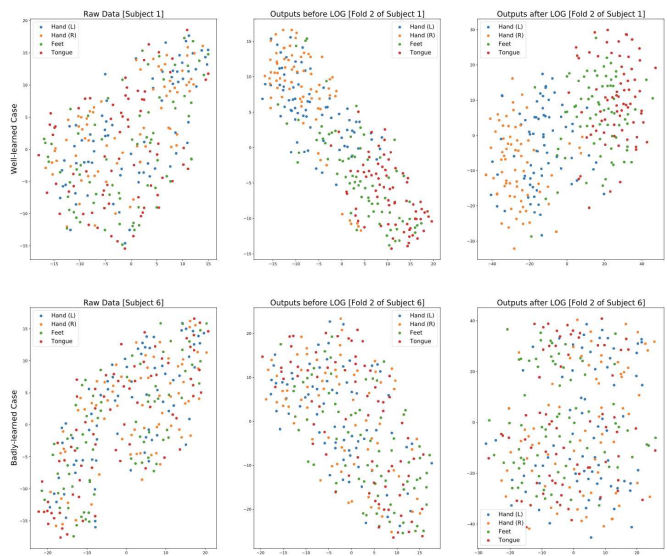}
\caption{Illustration of 2-dimensional projection for outputs of each intermediate stage in Tensor-CSPNet, including the \emph{well-learned} case (Subject No.1 in BCIC-IV-2a, Upper three Figures) and the \emph{badly-learned} case (Subject No.6 in BCIC-IV-2a, Bottom three Figures) using t-SNE.
Each subfigure has 288 points with four colors. 
1-CSPNet$^{(9,1,1)}$\_BN with $o=22$ achieves 0.864 and 0.483 average accuracy in two cases, respectively. 
\label{t-SNE}} 
\end{figure}

\begin{figure}[!h]
  \begin{subfigure}{0.5 \linewidth}
    \includegraphics[width = \linewidth]{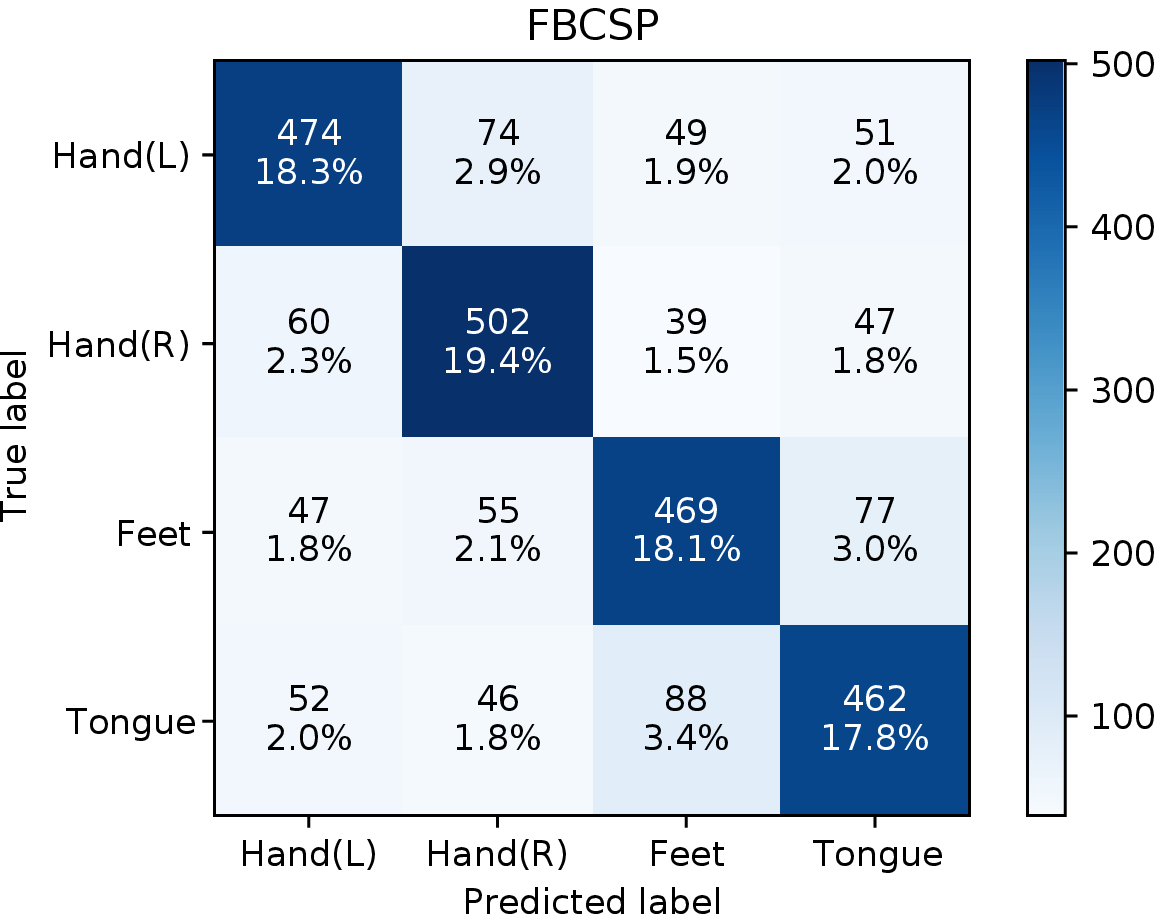}
  \end{subfigure}%
  \begin{subfigure}{0.5\linewidth}
    \includegraphics[width = \linewidth]{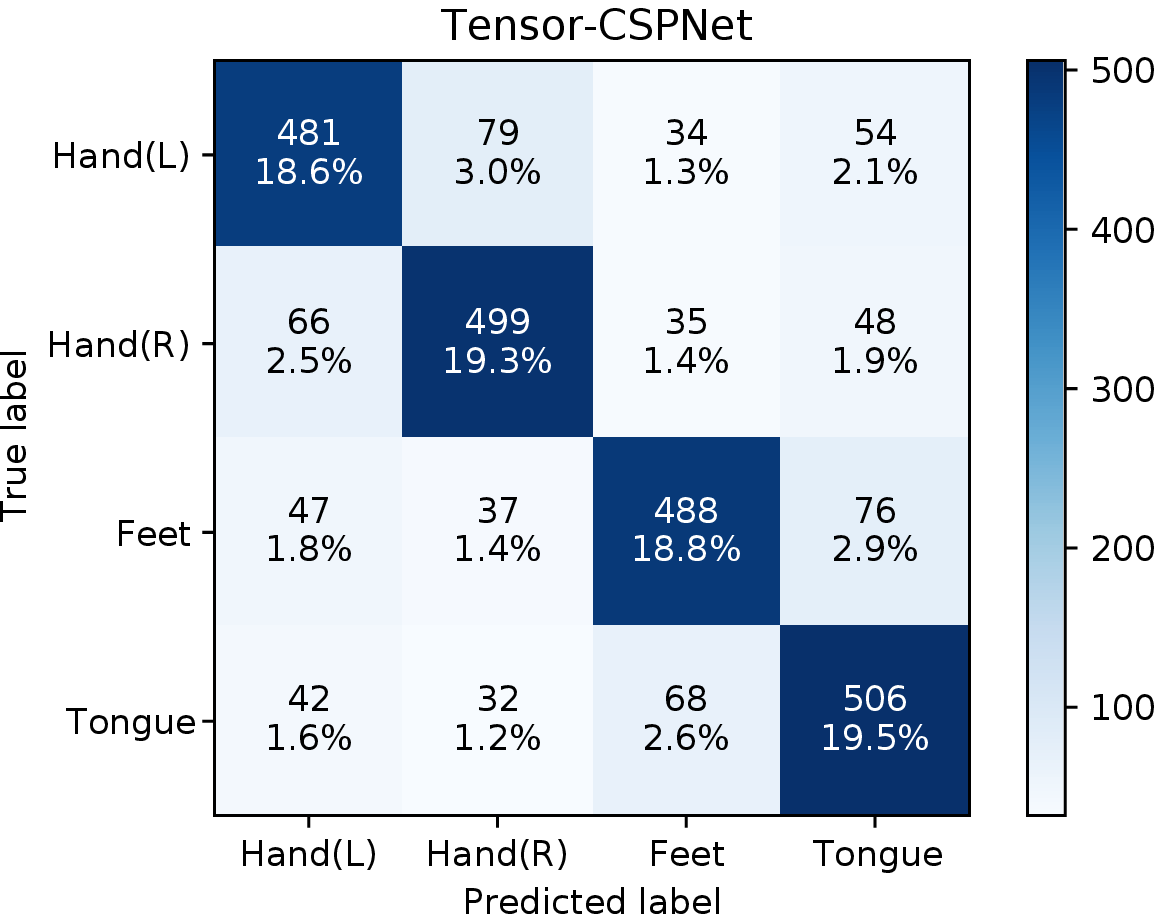}
  \end{subfigure}
  \caption{Unnormalized Confusion Matrices for FBCSP and Tensor-CSPNet in the CV scenario of BCIC-IV-2a:
4-class labels include Hand (L), Hand (R), Feet, and Tongue.
The total number of trials for both matrices is 2610 ( = 29 trails in test dataset $\times$ 10 folds $\times$ 9 subjects).
The quantities in each cell include the number of the corresponding class (top) and its percentage (\%) over the total number of trails (bottom). 
\label{fig:confusion_matrix}
}
\end{figure}

We investigate the visualization of BCIC-IV-2a using t-SNE. 
The \emph{well-learned} and \emph{badly-learned} cases are considered in the comparison, where the \emph{well-learned} and \emph{badly-learned} cases are with the average accuracy over 0.80 or under 0.50 under both Tensor-CSPNet and the FBCSP, respectively.  
The first line of Figure~\ref{t-SNE} is the \emph{well-learned} case upon Fold 2 of Subject 1.
The left subplot is the projection from the original data set.
We concatenate $22 \times 22$ SMCs derived from 9 band-pass signals as a $9\times 22 \times 22$-dimensional vector and project them via the t-SNE method.
The four colors in the resulting cross together, and the middle subplot is the projection from the outputs after the depth BiMap, Riemannian BN, and ReEig.
We notice that the four-color points are more separated than those in the left subplot and stacked in a sequence where the same color points concentrate on a specific location in the figure.
The right subplot is the projection from the outputs after LOG, and it is more concentrated for each color and, after that, more accessible for classification.
However, we observe that the configuration of the proposed approach is still hard to distinguish between two pairs, such as the pair of the Tongue (red) and the Feet (green) and the pair of Hand (L) and Hand (R). 
It is consistent with the statistics in the confusion matrix of Table~\ref{fig:confusion_matrix} that both FN (False-Negative) and FP (False-positive) for the tongue and feet are 2.6\% and 2.9\%, respectively. Both FN and FP for the hand (L) (blue) and hand (R) (yellow) are 2.5\% and 3.0\%, respectively.
The second line of the three subplots is the projections on the \emph{badly-learned} case.
All color points cross together from the first figure to the last one, and there is no clear statistical pattern in shape.
%

\subsection{Computational Efficiency}
This subsection investigates the computational efficiency of the Tensor-CSPNet. 
The experiments are conducted on an Intel (R) Xeon (R) CPU $@$ 2.20 GHz with one socket, two cores per socket, and two threads per core. 
There are two subtopics to discuss in this subsection, including the operation time per iteration and calibration time per subject. 

\subsubsection{Operation Time per Iteration}
\begin{figure}[!h]
\centering
\includegraphics[width=4.5cm]{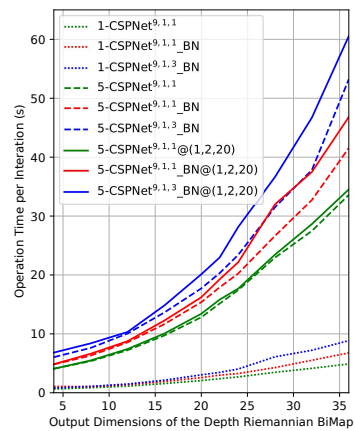}
\caption{Illustration of the relation between the operation time per iteration and the output dimension of the depthwise BiMap layer.\label{runtime_output_dimension}} 
\end{figure}

We pick three specific groups of different configurations of Tensor-CSPNet in Figure~\ref{runtime_output_dimension}.
Each group has three curves, and one color represents one group. 
The operation time per iteration increases as the output dimension $o$ of the depthwise BiMap layer increases.
The architecture without Riemannian BN (green) is the shortest operation time per iteration, the one with Riemannian BN (red) is in second place, and the one with both the Riemannian BN and a multi-layer classification stage (blue) is the longest.
In addition, the operation time per iteration of the group of 5-CSPNet rapidly increases $6\sim8$ times that of the group of 1-CSPNet.
The operation time per iteration of the group of 5-CSPNet$^{(9, 1, 1)}$\_{BN}@(1,2,20) is slightly longer than the ones of the group of 5-CSPNet.

\begin{table*}[!t]
\caption{Table for Architecture of $1$-CSPNet$^{(9, 1, 1)}$ with $o = 22$ on BCIC-IV-2a.}
\centering 
\begin{tabular}{l l r r }
\toprule
Stage &Layer   & Output Shape  & Parameters\\
\midrule
Tensor Stacking& 	& 	$ 9 \times 22 \times 22$ & /\\
			&Depthwise BiMap  & $9 \times 22 \times 22$  &  $9 \times 22 \times 22$ \\
CSP & Riemannian BN & $ 9 \times 22 \times 22$     &   $1 \times 22 \times 22$ \\
		& ReEig & $9 \times 22 \times 22$ & / \\
		& LOG  & $ 9 \times 22 \times 22$ & / \\
Classification &Linear Network & 4 & $4 \times 9 \times 22 \times 22$ \\
\midrule
Total &  & & 27,104 \\
\bottomrule
\end{tabular}
\label{tab:para_architecture_1}
\end{table*}

\begin{table*}[!t]
\caption{Table for Architecture of $5$-CSPNet$^{(9, 3, 1)}$ @(9, 5,10) with $o_1, o_2, o_3 = 22$ on BCIC-IV-2a.}
\centering 
\begin{tabular}{l l r r }
\toprule
Stage &Layer   & Output Shape  & Parameters\\
\midrule
Tensor Stacking& 	& 	$5\times 9 \times 22 \times 22$ & /\\
			&Depthwise BiMap  & $5\times 9 \times 22 \times 22$  &  $9 \times 22 \times 22$ \\
CSP (1st)& Riemannian BN & $5\times 9 \times 22 \times 22$     &   $1 \times 22 \times 22$ \\
		& ReEig & $5\times 9 \times 22 \times 22$ & / \\
		& LOG  & $5\times 9 \times 22 \times 22$ & / \\

		&Depthwise BiMap  & $5\times 9 \times 22 \times 22$ &  $9 \times 22 \times 22$ \\
CSP (2nd)&Riemannian BN & $5\times 9 \times 22 \times 22$ & $1 \times 22 \times 22$ \\
		&ReEig & $5\times 9 \times 22 \times 22$ & / \\
		&LOG  & $5\times 9 \times 22 \times 22$ & / \\

		&Depthwise  BiMap  & $5\times 9 \times 22 \times 22$  & $9 \times 22 \times 22$ \\
CSP (3rd)&Riemannian BN & $5\times 9 \times 22 \times 22$ & $1 \times 22 \times 22$ \\
		&ReEig & $5\times 9 \times 22 \times 22$ & / \\
		&LOG  & $5\times 9 \times 22 \times 22$ & / \\
Temporal Convolutional & 2D CNN & $(9-9+1) \times (5-5+1) \times 10$ & $10 \times 5 \times 9 \times 22 \times 22$ \\
Classification &Linear Network & 4 & $4 \times 10$ \\
\midrule
Total &  & & 232,360 \\
\bottomrule
\end{tabular}
\label{tab:para_architecture_2}
\end{table*}


\begin{table}[!h]
\caption{Experiments on Calibration Time per Subject: 
Average accuracy (Acc.) and standard deviation (Std.) for the calibration time (s) per subject of baselines and Tensor-CSPNet on BCIC-IV-2a. 
The output dimension of the CSP stage is $22$ for the three configurations of Tensor-CSPNet.}
\centering 
\begin{tabular}{ll r}
\toprule
{} & Avg. Times & Approx. \\ 
\toprule
MDM       & 53.24 sec&/\\
FBCSP    & 55.88 sec& /\\ 
TSM        & 60.45 sec& /\\
\midrule
FBCNet & 353.14 sec& 5.89 min \\
EEGNet & 500.03 sec&  8.33 min \\
SPDNet & 593.32 sec& 9.89 min \\
ConvNet & 1542.61 sec& 25.71 min  \\
\midrule
1-CSPNet$^{(9, 1, 1)}$\_{BN} &  1997.23 sec& 33.29 min \\
5-CSPNet$^{(9, 1, 1)}$@(1,2,20)  & 7650.99 sec&  2.13 hr \\
5-CSPNet$^{(9, 1, 1)}$\_{BN}@(1,2,20) & 8860.63 sec&  2.46 hr \\
\bottomrule
\end{tabular}
\label{tab:converge}
\end{table}

\subsubsection{Calibration Time per Subject}\label{duration}
Many training parameters in the DL approaches will affect the calibration time, such as learning rate, batch size, epochs, etc.
We fix batch size equivalent to the test set size 28 for training and validation, and the initial learning rate is 0.01 with decay. 
The total number of training epochs is default 60, and an early stopping strategy with 15 patience is adopted.
The average accuracy and standard deviation of the calibration time are shown in Table~\ref{tab:converge}. 
The average calibration time per subject for non-DL approaches is around 1 min.
However, the calibration time per subject for the DL approaches is much longer, especially 5-CSPNet$^{(9, 1, 1)}$\_{BN}@(1,2,20) is the longest.
Three configurations in Tensor-CSPNet are adopted for comparison. 
Firstly, 1-CSPNet$^{(9, 1, 1)}$\_{BN} is the simplest without temporal segmentation, and it has a longer calibration time than other DL approaches.
Because 9-time input size and bigger size of architecture extend the calibration time. 
Secondly, for 5-CSPNet$^{(9, 1, 1)}$@(1,2,20), the input size of both algorithms is five times that of 1-CSPNet$^{(9, 1, 1)}$\_{BN}.
Hence, there is no doubt that both have over two hours of calibration time.

\begin{table}[!h]
\caption{Table of number of parameters in network architecture.~\label{tab:number}
}
\centering 
\begin{tabular}{l  r  r }
\toprule
 {} & Shallow  & Deep\\
\midrule
EEGNet & 796 & 1,716  \\
ConvNet & 40,644 & 152,219  \\
Tensor-CSPNet & 27,104 & 232,360 \\
\bottomrule
\end{tabular}
\end{table}

\subsection{Architecture of Tensor-CSPNet}

We provide the detailed architecture of $1$-CSPNet$^{(9, 1, 1)}$ with $o=22$ and $5$-CSPNet$^{(9, 3, 1)}$ $@(1,2,20)$ with $o_1, o_2, o_3 = 22$ on BCIC-IV-2a.
Table~\ref{tab:para_architecture_1} and~\ref{tab:para_architecture_2} exhibit the total parameters of two configurations are 27104 and 232360, respectively.
Table~\ref{tab:number} exhibits the numbers of parameters of different architecture as follows,
The numbers of ConvNet and EEGNet refers to Table~{\color{blue} 3}~\cite{lawhern2018eegnet}. 
From Table~\ref{tab:number}, we notice that EEGNet has a tiny size of parameters. 
In contrast with ConvNet and EEGNet, the architecture of Tensor-CSPNet needs large-scale parameters to preserve the geometric information of SCMs.


\end{document}